\documentclass[9pt,twocolumn,twoside]{extarticle}
\usepackage[margin=0.75in]{geometry}
\usepackage{amsmath,amssymb}
\usepackage{authblk}
\usepackage{booktabs}
\usepackage{graphicx}
\usepackage[numbers,sort&compress]{natbib}
\usepackage[colorlinks=true,allcolors=blue]{hyperref}
\usepackage{mdframed}
\usepackage{comment}
\usepackage{color}
\usepackage{xcolor}
\usepackage{soul}
\usepackage{xurl}
\usepackage{subcaption}

\usepackage{placeins}
\newcommand{\authorcontributions}[1]{\def\storedauthorcontributions{#1}}
\newcommand{\authordeclaration}[1]{\def\storeddeclaration{#1}}
\newcommand{\correspondingauthor}[1]{\def\storedcorrespondingauthor{#1}}
\newcommand{\keywords}[1]{\def\storedkeywords{#1}}

\newenvironment{resultbox}
  {\begin{mdframed}[backgroundcolor=gray!5,linecolor=black!40,
    linewidth=0.5pt,roundcorner=2pt,innerleftmargin=5pt,
    innerrightmargin=5pt,innertopmargin=5pt,innerbottommargin=5pt]}
  {\end{mdframed}}

\begin{document}
\onecolumn


\title{Low-vocality engagement and participation dynamics on Bluesky}

\author[a,b]{Veronica Mesina}
\author[a,b]{Andrea Failla}
\author[b,c]{Luca Pappalardo}
\author[b]{Giulio Rossetti}

\affil[a]{Department of Computer Science, University of Pisa, Pisa, Italy}
\affil[b]{Institute of Information Science and Technologies "A. Faedo" (ISTI), National Research Council (CNR), Pisa, Italy}
\affil[c]{Scuola Normale Superiore, Pisa, Italy}

\authorcontributions{V.M, A.F., and G.R. designed research;
V.M, A.F., L.P., and G.R. performed research; 
A.F. collected the data; 
V.M. and A.F. analyzed the data; 
V.M. and A.F. wrote the paper. All authors read and approved the final version of the manuscript.}
\authordeclaration{The authors declare no conflict of interest.}

\correspondingauthor{veronica.mesina@phd.unipi.it}


\maketitle

\begin{mdframed}[backgroundcolor=blue!4,linecolor=blue!40,
  linewidth=0.5pt,innerleftmargin=7pt,innerrightmargin=7pt,
  innertopmargin=6pt,innerbottommargin=6pt,nobreak=true]
\noindent This is a preprint of an article published in \emph{EPJ Data Science}. The final authenticated version is available online at: \url{https://doi.org/10.1140/epjds/s13688-026-00711-9}
\end{mdframed}

\begin{abstract}
Online participation is often measured through visible expression, especially posting, yet many consequential forms of engagement occur through less vocal actions such as liking and following. Here we study how users inhabit Bluesky by reconstructing participation profiles from more than three billion activity records produced by a near-complete sample accounting for more than 80\% of registered users. We aggregate behavior into monthly user-level observations and distinguish two dimensions that are often conflated in platform analytics: intensity, capturing how much users engage, and style, capturing how engagement is expressed across actions. We find that vocal production is highly concentrated, but low-posting behavior does not imply absence from platform participation. High-intensity engagement is most strongly associated with liking rather than posting, while posting-oriented participation is more common among low-intensity users, indicating that visibility and sustained engagement should not be conflated. Transition patterns suggest that high-intensity likers and posters could be described as attractors; network-building redirects users within the active space; whereas observed inactivity acts as a persistent boundary that selectively limits re-entry. Higher-order motifs further show that inactivity often interrupts rather than erases prior regimes, and that low-intensity liking can precede durable high-intensity engagement. These results show that online participation is structured by differentiated low-vocality practices, calling for a shift from post-centered measures of activity toward dynamic accounts of platform presence. We identify a broader challenge for computational social science: platform participation cannot be adequately understood through the behavior of vocal minorities alone. Because the most visible traces are not necessarily those most associated with sustained engagement, incorporating low-vocality forms of participation can help counter visibility biases embedded in platform data and provide a more representative account of how platform life is sustained.
\end{abstract}

\begin{mdframed}[backgroundcolor=gray!4,linecolor=black!30,
  linewidth=0.5pt,innerleftmargin=7pt,innerrightmargin=7pt,
  innertopmargin=6pt,innerbottommargin=6pt,nobreak=true]
\noindent\textbf{Keywords}\par
\storedkeywords

\medskip
\noindent\textbf{Author contributions}\par
\storedauthorcontributions

\medskip
\noindent\textbf{Competing interests}\par
\storeddeclaration

\medskip
\noindent\textbf{Corresponding author}\par
\storedcorrespondingauthor
\end{mdframed}

\clearpage
\twocolumn
\section*{Introduction}
Online Social Platforms (OSPs) constitute an unprecedented observational social infrastructure~\cite{van2018platform, di2026patterns, Sajadi_2018}, providing behavioral traces from online social life at a level of scale and granularity that would have been impossible through conventional methods alone~\cite{pera2026measuring,milli2025engagement,desiderio2025highly}. Nevertheless, the most visible traces are not necessarily the most representative of participation: they are produced by users who post, comment, and publicly express themselves, while much of what sustains and transforms platform environments unfolds through quieter, lighter, and less discursive actions such as reading, liking, following, amplifying~\cite{boyd2010social, Marwick_boyd_2011, guan2025using, crawford2011listening}. This constitutes a structural tension at the heart of computational social science: platform data appear abundant, but their abundance is uneven, so that what is most readily measurable is not necessarily what is most socially representative~\cite{oswald2025tip, hargittai2020potential, bright2020power}. At the same time, this uneven observability matters sociologically because platforms are not merely repositories of behavioral traces, but digital spaces in which social life is organized through platform-specific practices, affordances, and forms of visibility~\cite{pariser2011filter, boyd2010social, bucher2018affordances, treem2013social, magaudda2015bourdieu}.

\indent This structural asymmetry between visible and less visible participation has been conceptualized through a set of recurring analytical vocabularies --- engagement inequalities~\cite{baqir2023beyond, baqir2025unveiling}, the lurker/active user distinction~\cite{nonnecke2000lurker, nonnecke2001lurkers, nonnecke2003silent}, and the silent majority versus vocal minority~\cite{mustafaraj2011vocal, zhou2024unveiling, gearhart2015something, noelle1974spiral}. Each of these has proven essential for naming a fundamental feature of online social life: visible contribution is highly concentrated, with a small minority producing a disproportionate share of public content while larger populations read, observe, and/or contribute only intermittently. 

\begin{figure*}[t!]
\centering

\includegraphics[width=\linewidth]{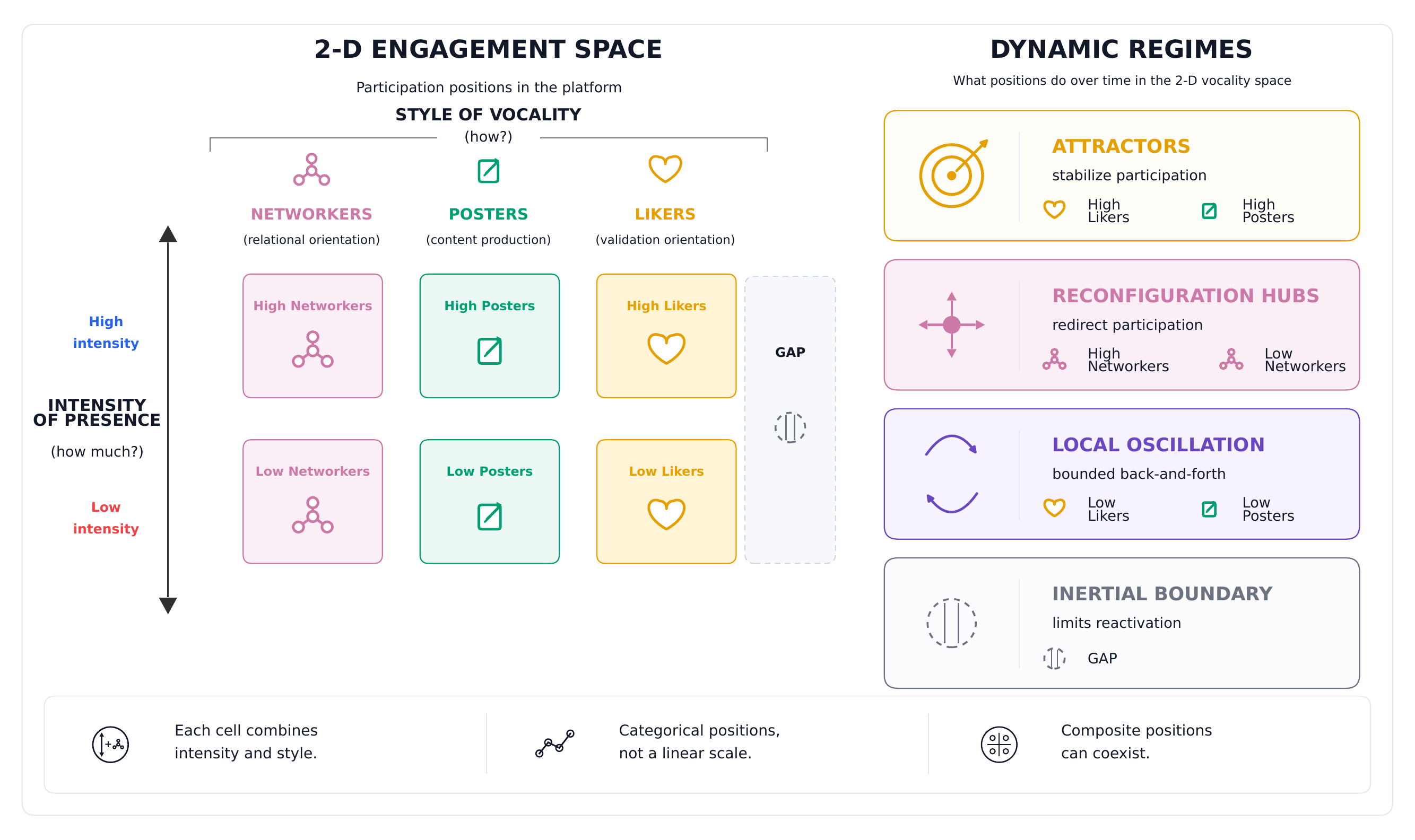}

\caption{User participation is represented along two empirically derived dimensions: \emph{intensity of presence} (how much users engage with the platform) and \emph{style of vocality} (how engagement is expressed across actions). Both dimensions are obtained through unsupervised clustering of behavioral features, yielding two intensity profiles (High and Low Intensity) and three style profiles (Likers, Posters, and Networkers). Their combination defines six data-driven participation positions as shown in \ref{fig:stories}, while \emph{Gap} denotes periods of observed inactivity. The right panel summarizes four dynamic regimes identified from transitions between these positions over time: \emph{Attractors}, which stabilize participation; \emph{Reconfiguration Hubs}, which redirect users across active positions; \emph{Local Oscillations}, which capture bounded movement between related low-intensity profiles; and the \emph{Inertial Boundary}, which limits reactivation following inactivity. Together, these components conceptualize participation as an empirically reconstructed engagement space rather than a predefined typology or a linear continuum of engagement.}
\label{fig:abstract}
\end{figure*}

\noindent Whether formalized through the 90–9–1 rule and the 1\% rule~\cite{van20141, antelmi2019characterizing, carron2014describing}, the broader Pareto-like 80/20 logic~\cite{liu2020does, matei2015pareto}, or the Spiral of Silence framework~\cite{noelle1974spiral, zhou2024unveiling, haug2025supporting, gearhart2015something}, these formulas are best understood as heuristics pointing to the same structural asymmetry rather than universal laws, given that the empirical distribution of participation varies considerably across platforms, communities, and topics — with the proportion of \textit{silent users} fluctuating between 70\% and 90\%~\cite{wang2019silent, tagarelli2013s, gong2015characterizing}. 
Computational approaches have extended these frameworks, using a number of methods to render low-visibility user data more tractable: estimating latent opinions through network analysis~\cite{gong2015characterizing, wang2019silent, tagarelli2014lurking, tagarelli2013s} and agent-based models~\cite{alvarez2016network, aranda2024sound}, correcting for under-represented users~\cite{karami2023silence}, and incorporating implicit engagement signals into models of diffusion and stance formation~\cite{zhou2024unveiling, wang2019silent, cui2021voice, baqir2023beyond, baqir2025unveiling}. These contributions have opened an important methodological space of inquiry. However, operationalizing \textit{silence} or, at least, being \textit{less active}, within these frameworks comes at a cost: silent users are often treated as actors whose absent voice must be recovered or whose latent stance must be inferred~\cite{gong2015characterizing}; low-vocality participation tends to be compressed into a residual category defined primarily by what it lacks while the boundary between active and passive behavior remains too vaguely defined across studies. \\

The category of silence, in particular, is too broadly defined to capture the heterogeneity of low-vocality participation and risks foreclosing inquiry into its internal differentiation, collapsing heterogeneous practical and relational forms of engagement through which users contribute to platform life into an undifferentiated residual --- that is, silence or passivity --- thus distorting our understanding of the broader social space from which they emerge~\cite{hargittai2020potential, papacharissi2009virtual, adjin2023lurking}. Visible contribution should not be treated as a univocal indicator of sustained platform use: users may post, comment, or otherwise become publicly visible in episodic and/or intermittent ways, without maintaining a dense presence on the platform. 
By the same token, non-posting may correspond to very different practical orientations~\cite{adjin2023lurking, preece2004top, sun2014understanding, crawford2011listening, crawford2009following, giuffre2026youth}: learning community norms, avoiding conflict, protecting privacy, gathering information, or simply finding that reading already satisfies the user's purpose. 
Consuming content without producing it should not be thought of as a backstage against which participation is performed; in fact, it is itself part of the practical organization of digital sociality~\cite{papacharissi2009virtual, hargittai2007whose, magaudda2015bourdieu, Sajadi_2018}. 
The same logic applies to low-effort actions, which are not necessarily weak social actions.
A like contributes to visibility, feedback, and ranking; a follow reorganizes the user's relational and informational environment~\cite{GREEN_MCCABE_SHUGARS_CHWE_HORGAN_CAO_LAZER_2025}; a return after inactivity may reactivate a previously stabilized mode of presence without signaling renewed vocality. Low-vocality participation is structured by its own internal logic — one that calls for analytical frameworks capable of registering it on its own terms rather than as the absence of something more. Silence --- or, more precisely, what is less visible --- should therefore be understood not as a null condition, but as a differentiated set of practical orientations within the space of possible actions made available and constrained by platform affordances~\cite{boyd2010social, bucher2018affordances}. \\

This article approaches platform participation through a relational perspective. Rather than treating engagement as an individual-level property to be inferred from observable traces, it conceives participation as a space of differentiated practices~\cite{hargittai2007whose,magaudda2015bourdieu}, where users' positions are defined relationally through differences, distances, and trajectories~\cite{bourdieu1984distinction,bourdieu1989social,bourdieu1992invitation}. This frame makes it possible to construct a map of platform practices in which actions acquire meaning through their position relative to other actions and through the temporal paths that connect them. Empirically, we operationalize this space through two dimensions that are often conflated in platform analytics (Fig.~\ref{fig:abstract}). \textit{Intensity} captures how much users engage with the platform, including how often they are active, how many actions they perform, how much time they spend, and how dense their activity is within sessions. \textit{Style} captures how engagement is expressed across qualitatively different practices, including but not limited to liking, posting, reposting, and following (full action set is available in the Methods). Neither dimension is defined \textit{a priori}; both are recovered from behavioral traces through a data-driven procedure that allows recurrent configurations to emerge from the structure of the data.\\

We examine these dynamics on Bluesky, a decentralized social platform whose public user repositories make it possible to observe heterogeneous action traces at scale~\cite{kleppmann2024bluesky,failla2024m,quelle2025bluesky}. 
Moreover, previous research has shown log-based records a more direct basis for reconstructing observable usage patterns~\cite{parry2021systematic}. We therefore rely on Bluesky activity logs to derive behavioral measures of platform presence, particularly the intensity with which users engage over time. Starting from approximately three billion activity records produced by 81\% of registered users, we reconstruct monthly user-level observations over an 18-month window from July 2023 to December 2024. The user-month is the unit of analysis: it allows us to characterize how users participate during each period and how they move across participation profiles over time. To reduce boundary-related artifacts, we exclude the earliest phase of platform activity and avoid right-censored observations at the end of the collection period. 
The analysis proceeds in four steps. First, we reconstruct activity sessions from timestamped actions and derive monthly user-level features. Second, we cluster observations separately in the intensity and style feature spaces, obtaining interpretable profiles of how much users engage and how they distribute their activity across practices. Third, we combine these profiles into two-dimensional intensity--style configurations. Fourth, we analyze movement across configurations using transition matrices, a within-user temporal-randomization null model, and higher-order sequential motifs~\cite{pei2004mining}. This design allows us to distinguish persistence and attraction areas from transitions that are inhibited relative to the null model, and to identify recurrent multi-period trajectories such as interruption, return, and escalation. Further details on data collection, sessionization, clustering, transition analysis, and pattern mining are provided in Materials and Methods.\\

The results show that vocal production on Bluesky is highly concentrated, but low-posting behavior does not imply absence from platform participation. High-intensity engagement is most strongly associated with liking rather than posting, indicating that sustained participation is often organized around lightweight validation. Network-building functions as a mechanism of reconfiguration within the active space, while the Gap, our operational measure of observed inactivity, operates as an inertial boundary that selectively limits re-entry. These findings show that low-vocality practices are not marginal residues of participation, but consequential modes through which users sustain, redirect, and resume platform presence. 

\section*{Results}
\begin{figure*}[t!]
\centering
\includegraphics[scale=0.5]{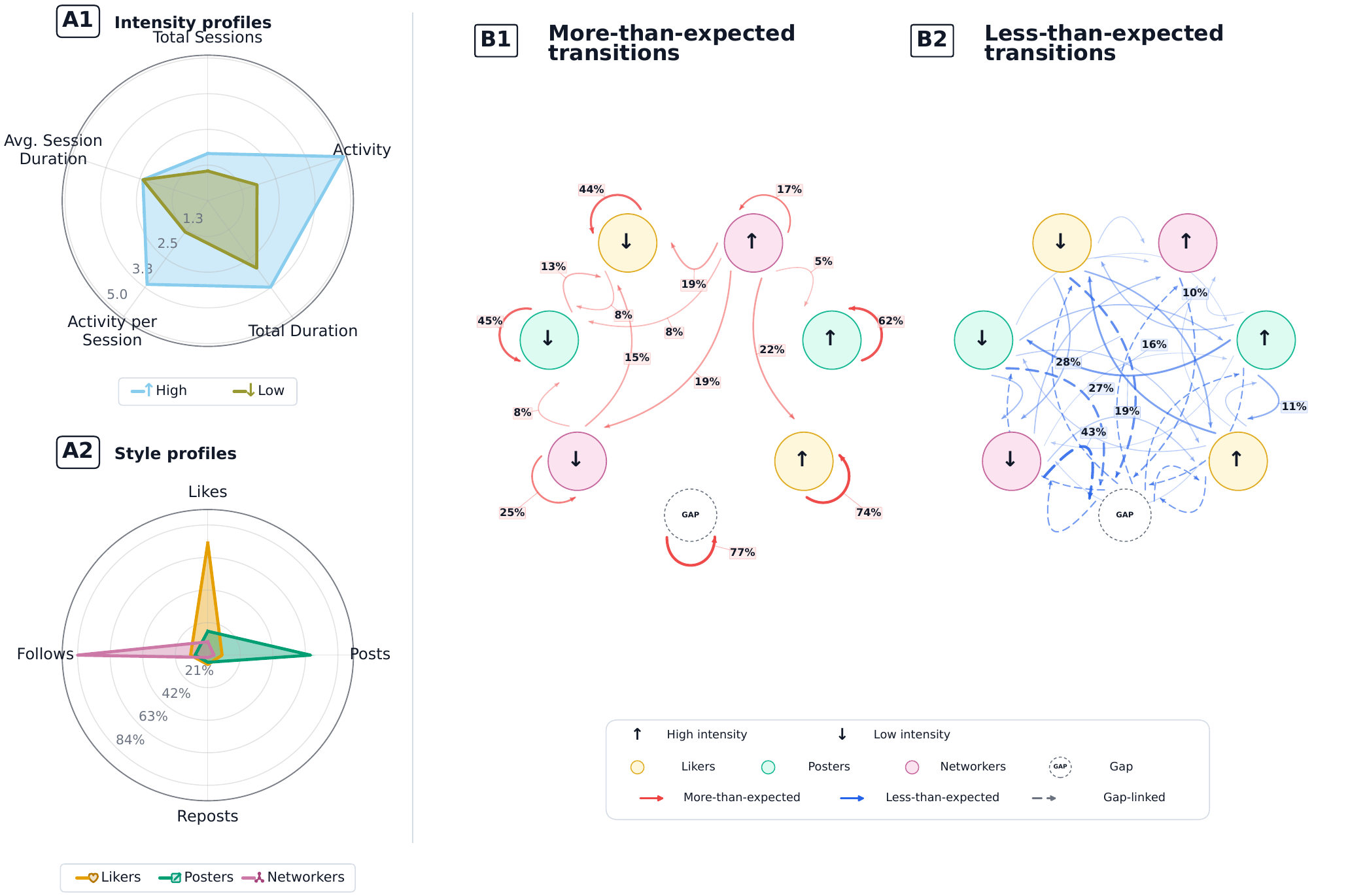}
\caption{
Overview of participation profiles and transition dynamics.
Panel A reports the cluster centroids for the intensity and style feature spaces. In Panel A2, only style features with centroid values of at least 0.05 are shown to reduce visual clutter; the complete feature set is reported in the Materials and Methods section.
Panel B represents transitions between two-dimensional intensity--style profiles. Edge labels report the observed transition probability $P_{\mathrm{obs}}$; red and blue edges indicate transitions that are over- and under-represented relative to the null model, respectively.
We find that participation on the platform is organized into distinct and interpretable user profiles, and that these profiles display temporal stability and statistically non-random transition structure.
}
\label{fig:main}
\end{figure*}

We first establish a post-centered baseline for participation. Treating vocality as the total number of posts authored by each user, we find that 65.37\% of users authored at least one post during the observation window. At the same time, vocal production was highly concentrated. The top 20\% of users produced 95.88\% of all posts, and only 240,956 users, corresponding to 5.77\% of the observed population, accounted for 80\% of the 448M posts in the dataset.
Thus, Bluesky does not display a silent majority in the strict sense of non-posting users. Rather, it displays a strong concentration of vocal production, in which posting activity is dominated by a small fraction of highly vocal users. This baseline captures an important asymmetry, but it also illustrates the limits of a post-centered account of participation. Users who post rarely or not at all may still participate by liking, reposting, following, returning after inactivity, or reorganizing their informational environment. We therefore treat vocality as only one observable dimension of platform participation and reconstruct a broader behavioral space defined by intensity, style, and temporal trajectories.\\
In the following, we characterize and interpret user typology with respect to the identified behavioral profiles. 
Average feature values for each group are displayed in Figure~\ref{fig:main}A, while support counts are reported in Table~\ref{tab:cluster_summary}.

\begin{table}[t]
\centering
\scriptsize
\caption{Summary statistics for clusters in the \textit{Intensity} and \textit{Style} feature spaces. Observations denote user-month observations assigned to each cluster; percentages are relative to the total user-month observations in the corresponding dataset. Unique users denote distinct users appearing at least once in a cluster. Users observed multiple times in the same cluster are counted once, while users appearing in different clusters across months are counted once in each relevant cluster. Thus, cluster-specific unique-user counts are not additive across clusters.}
\label{tab:cluster_summary}
\begin{tabular}{@{}lcc@{}}
\toprule
Cluster & \# Observations & \# Unique users \\
\midrule
High-Intensity & 7\,303\,384 (35.3\%) & 1\,968\,590 \\
Low-Intensity  & 13\,351\,560 (64.6\%) & 3\,952\,558 \\
\addlinespace
Likers & 10\,151\,191 (49.1\%) & 2\,348\,656 \\
Posters  & 4\,684\,430 (22.6\%)  & 1\,569\,889 \\
Networkers  & 5\,819\,323 (28.1\%)  & 2\,806\,358 \\
\bottomrule
\end{tabular}
\end{table}

\FloatBarrier

\begin{resultbox}
\small
\textbf{Key results}: 
\begin{itemize}
    \item Post-centered measures reveal inequality in vocal production, but do not capture the full structure of platform participation. 
    \item Users separate into high- and low-intensity profiles and into distinct participation styles: liking, posting, and network-building.
    \item High-intensity participation is predominantly liking-oriented: \(66\%\) of high-intensity observations fall into the Liker profile.
    \item Low-intensity participation is more heterogeneous, showing that posting and network-building can occur without sustained high-intensity engagement.
\end{itemize}
\end{resultbox}


\subsubsection*{Intensity}
With respect to usage intensity, we identify a multidimensional boundary separating high- and low-activity users. 
The main distinction does not lie simply in the total time spent on the platform, but rather in interaction density, measured as the volume of actions relative to the number of sessions. 
The \textit{High-Intensity} cluster accounts for 36\% of user-month observations. 
It is characterized by systematically high values across all intensity-related features, with particularly high levels of total activity and activity per session. Users in this group access Bluesky frequently, perform many actions overall, and remain highly active within each session. 
\\ \ \\
The \textit{Low-Intensity} cluster accounts for 64\% of user-month observations. 
By contrast, it displays substantially lower values across almost all features, with similar average session duration. 
This configuration points to a more observational mode of use, characterized by extended presence on the platform but limited interaction. 
This pattern suggests a form of participation oriented toward browsing, reading, and scrolling, that is, toward a more passive consumption of content rather than direct interaction with the interface or with other users.

\subsubsection*{Style}
With respect to engagement style, we identify three clearly differentiated profiles defined by a distinct distribution of activity types. 
Specifically, we note a sharp concentration of actions within each cluster, suggesting that participation is organized around highly specialized modes of engagement. 
The three clusters thus reflect distinct ways of organizing and expressing participation, with each profile capturing a specific functional role within the platform.
\\ \ \\
The \textit{Liker} profile accounts for 50\% of user--month observations. 
It is characterized by a clear predominance of likes, which account for about 72.7\% of actions, followed by smaller shares of new follows (11.0\%) and published posts (9.1\%). 
The participation style of this type of user is primarily articulated through feedback and endorsement rather than original content production or network-building activities. 
Despite limited content production, these users play a key role because they sustain everyday engagement and shape visibility through frequent, low-cost signals of attention and approval.
\\ \ \\
The \textit{Poster} profile accounts for 22\% of user--month observations. 
It is dominated by published posts, which represent 66.4\% of actions, whereas likes account for a secondary share (15.7\%). 
This cluster defines a participation style centered on expression and production. 
Although smaller in size, this profile is structurally relevant because it contributes a substantial share of the content on which the other participation styles also rely. 
More broadly, it captures a generative mode of engagement, oriented less toward endorsement or exploration than toward the active production of material for the platform ecosystem.
\\ \ \\
The \textit{Networker} profile accounts for 27\% of user--month observations. 
It is defined by a very high concentration of follow actions, which account for about 83.9\% of practices, together with a residual share of likes (8.8\%). 
This cluster describes a profile that engages primarily in expanding and reorganizing the social graph, rather than in content production. 
This configuration suggests a participation style more focused on constructing one’s informational environment through connection choices and may also reflect phases in which users invest primarily in building their feed and selecting information sources.

\begin{figure*}[t!]
\centering
\includegraphics[scale=0.33]{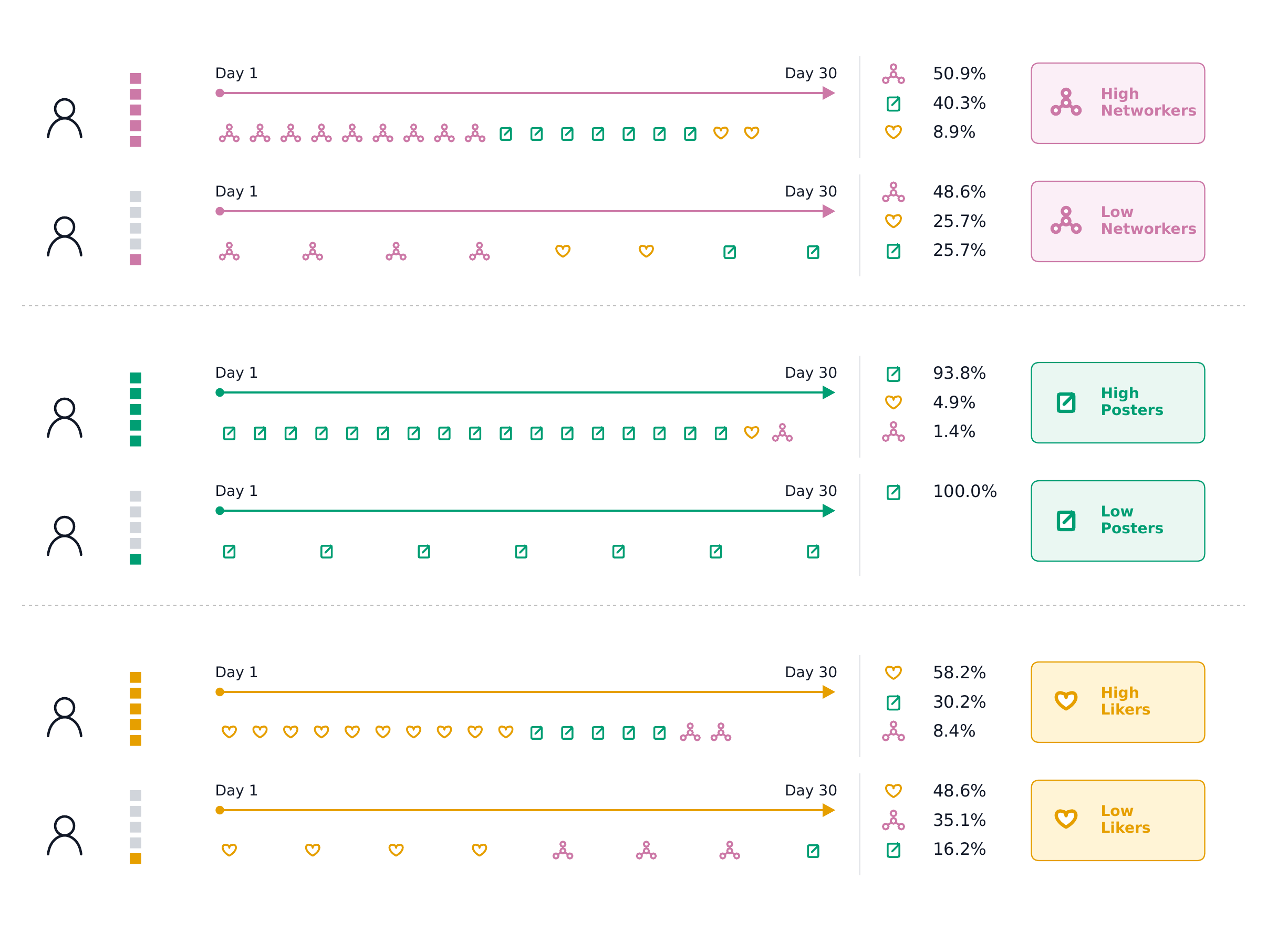}

\caption{The figure reports six representative user-month observations, one for each participation position obtained by crossing the intensity and style clusters. These composite states provide a more comprehensive representation of participation by jointly capturing both the level of engagement and the predominant style through which activity is expressed. As empirically derived configurations emerging from behavioral clustering, they illustrate how distinct forms of platform participation can be interpreted as positions within the 2-D representation of platform’s digital social space.}
\label{fig:stories}
\end{figure*}

\subsection*{Characterizing Two-dimensional Engagement Profiles}

\begin{figure*}[!tbp]
\centering
\includegraphics[width=0.75\linewidth]{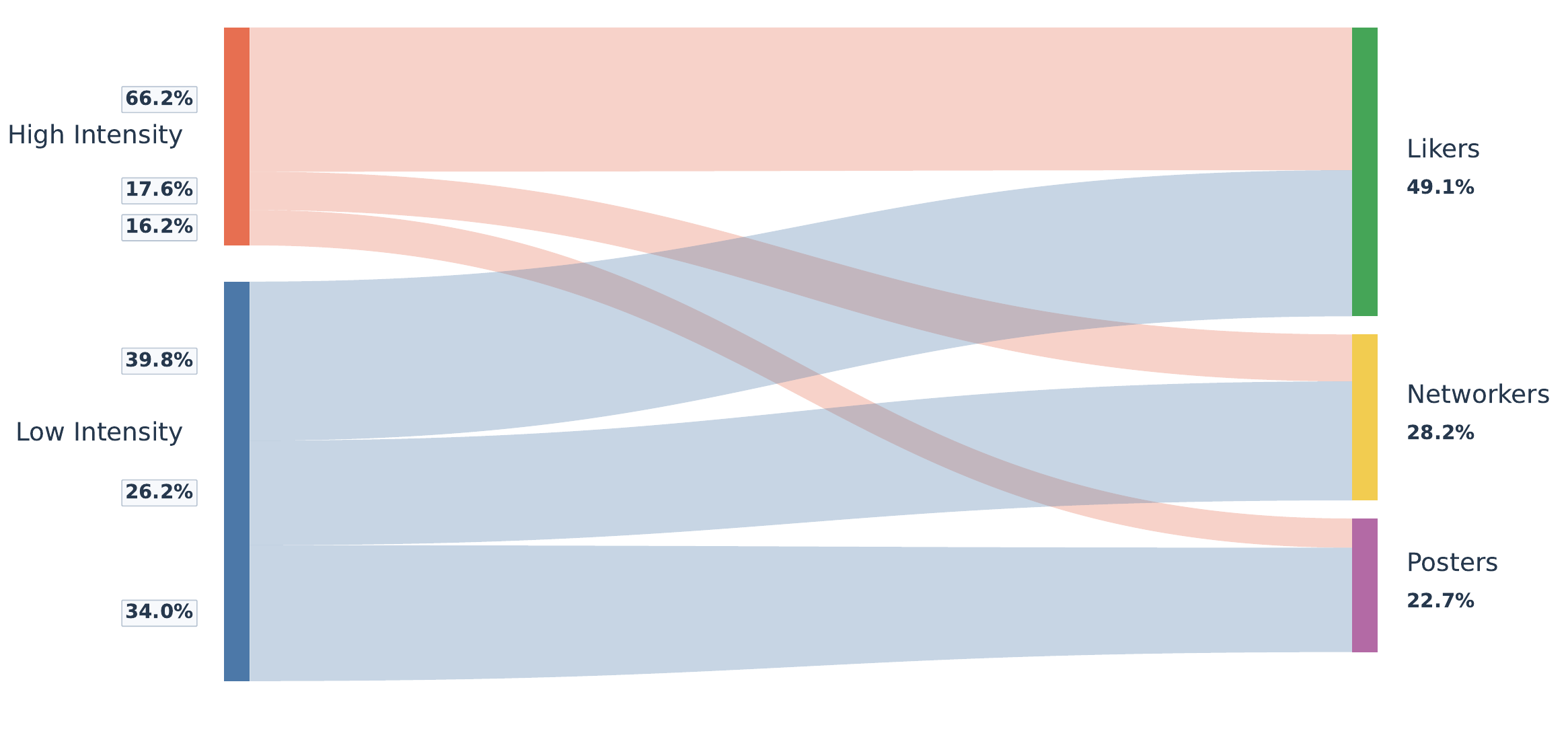}
\caption{
Distribution of two-dimensional participation profiles obtained by crossing intensity and style clusters.
Flows connect each intensity profile to the corresponding style profile across user--month observations. 
Flow thickness indicates the relative frequency of each intensity--style combination over the full observation period. 
Percentages on the left are computed within each intensity cluster, while percentages on the right report each style cluster's share of the total distribution. 
The diagram describes the composition of participation profiles and does not imply temporal ordering.
}
\label{fig:sankey_log_share}
\end{figure*}

Users with similar levels of intensity can enact different forms of participation, while the same engagement style can occur at different levels of activity, as shown in Figure~\ref{fig:stories}, where a small data sample of each possible configuration is presented. 
Therefore, we next move into the characterization of the six two-dimensional profiles that jointly capture \textit{how much} and \textit{how} users participate.

\noindent The co-occurrence analysis reported in Figure~\ref{fig:sankey_log_share} reveals that intensity and style are unevenly associated. 
The strongest association is between \textit{High-Intensity} and \textit{Likers}: 66\% of high-intensity observations fall into the liker profile.
Sustained participation is therefore not primarily associated with posting or network-building, but with recurrent forms of lightweight validation.
The \textit{Liker} profile also receives a substantial contribution from low-intensity users, almost 40\% of whom fall into this style, making it the dominant style overall.\\

\noindent By contrast, \textit{Low-Intensity users} are more broadly distributed across styles.
Compared with \textit{High-intensity users}, they are more strongly represented among \textit{Networkers} and \textit{Posters}, which account for 34\% and 26.2\% of low-intensity observations, respectively.
The higher share of \textit{Posters} among low-intensity users indicates that visible content production is not necessarily tied to the highest levels of overall activity, indicating that expressive participation can occur without sustained high-intensity engagement. This pattern is particularly notable when read alongside the predominance of Likers among high-intensity users. While high-intensity participation is primarily organized around recurrent acts of validation, posting-oriented users are relatively more common among low-intensity participants. Visible content production is therefore not necessarily tied to the highest levels of overall activity, suggesting that participation intensity and participation style capture partially distinct aspects of platform behavior.

\subsection*{User Profile Trajectories}
We next analyze whether users move between profiles in systematic ways over time.
For each user, we represent the monthly sequence of assigned $intensity \times style$ profiles as a trajectory and estimate transition probabilities between consecutive months.
To account for interruptions in observed activity, we introduce an additional state, \textit{Gap}.
This state identifies months in which a user is not observed as active within their trajectory.

For each pair of states, and with the Gap treated as a state of inactivity, the transition matrices report the difference between the observed transition probability and the average probability expected under a null model obtained by randomizing the temporal order of individual sequences across (N=100) independent realizations:

\begin{equation}
\Delta P = P_{\mathrm{obs}} - \langle P_{\mathrm{null}} \rangle .
\end{equation}

Positive values therefore indicate transitions that occur more often than expected under temporal independence, whereas negative values indicate transitions that are underrepresented relative to the null model.

Statistical significance is assessed by comparing the observed transition probabilities with those from the null model realizations, using a significance threshold of $\alpha = 0.01$. This criterion is used as a filtering condition for the transition matrices. In practice, however, all transitions satisfy the significance requirement.
An analysis of transitions for one-dimensional profiles is given in the SI.

\subsubsection*{Dynamic Engagement Regimes}

\begin{figure}[!htbp]
\centering
\includegraphics[width=\linewidth]{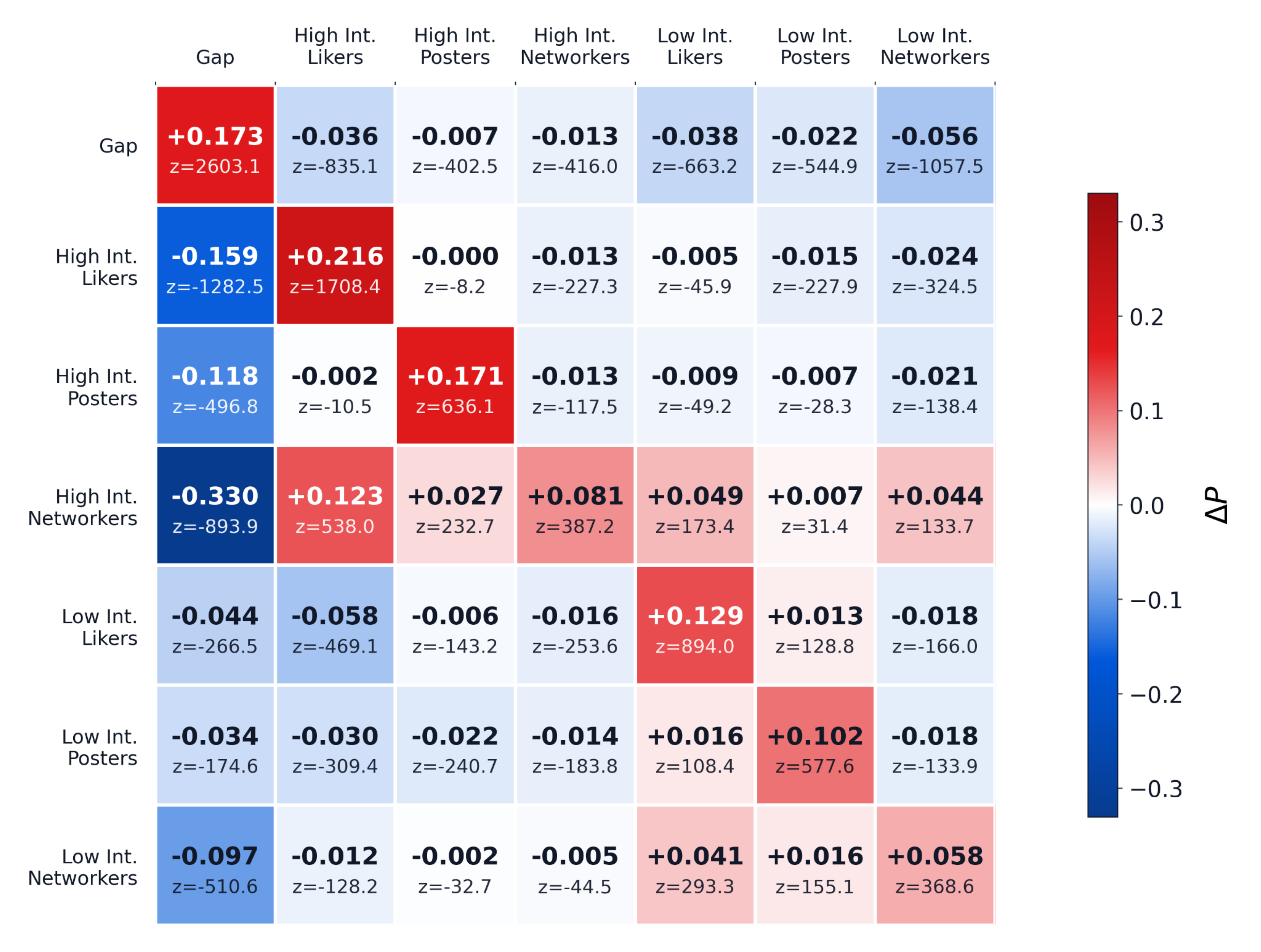}
\caption{2-D archetype transitions. Values in bold indicate $\Delta P = P_{obs} - P_{null}$, while the values below are the associated z-scores. Red and blue cells indicate stronger- and weaker-than-expected effects, respectively.}
\label{fig:configuration_transition_matrix}
\end{figure}

The joint transition matrix in Fig.~\ref{fig:configuration_transition_matrix} shows that movement across two-dimensional engagement configurations is highly structured. All transitions differ significantly from the temporal-randomization null model. The first general pattern is persistence: every configuration has a positive diagonal deviation, meaning that users remain in the same configuration more often than would be expected if the temporal order of their states were random. However, the off-diagonal structure shows that persistence is only part of the story. Different configurations also play different dynamic roles, depending on whether they stabilize participation, redistribute users across active states, support local movement between similar profiles, or inhibit re-entry after inactivity.

\begin{resultbox}
\small
\textbf{Key result}: The two-dimensional transitions reveals distinct dynamic regimes.
\begin{itemize}
    \item High-intensity Likers and Posters act as attractors, stabilizing active participation.
    \item High-intensity Networkers operate as a broad reconfiguration hub, whereas Low-intensity Networkers perform a more bounded, boundary-adjacent form of reconfiguration.
    \item Low-intensity Likers and Posters form a local oscillation zone, while the Gap acts as an inertial boundary that persists and selectively limits re-entry.
\end{itemize}
\end{resultbox}

We therefore interpret the transitions through four descriptive regimes. \emph{Attractors} are configurations with strong self-persistence and high diagonal transition probabilities. \emph{Reconfiguration hubs} are configurations that retain some persistence but also redistribute users toward multiple active states. \emph{Local oscillations} are reciprocal transitions between closely related configurations at the same intensity level. \emph{Inertial boundaries} are states that strongly reproduce themselves while limiting transitions into and out of active participation. These regimes indicate that participation is neither a linear progression from inactivity to activity nor a set of freely interchangeable behavioral positions. Instead, intensity and style combine into configurations with distinct temporal roles.

\paragraph*{Attractors}High-intensity Likers and High-intensity Posters are the main attractors of the active space. High-intensity Likers show the strongest self-persistence among active configurations, with \(\Delta P = +0.216\) and \(P = 0.74\). When users leave this state, they most often move to Low-intensity Likers, \(P = 0.16\), suggesting that intensity declines while engagement style is preserved. High-intensity Posters show a similar, though weaker, pattern, with \(\Delta P = +0.171\) and \(P = 0.62\). Their main outflow is toward Low-intensity Posters, \(P = 0.19\), followed by High-intensity Likers, \(P = 0.11\). These results indicate that high-intensity engagement is not homogeneous: it contains at least two stable regimes, one organized around routinized validation and the other around sustained content production.

\paragraph*{Reconfiguration Hub}
The networking configurations play different temporal roles depending on intensity. High-intensity Networkers operate as the clearest reconfiguration hub within the active space. They retain moderate self-persistence, $\Delta P = +0.081$, but also redirect users toward other active configurations, most notably High-intensity Likers, $P = 0.22$ and $\Delta P = +0.123$. At the same time, their transition to the Gap is both relatively rare and strongly underrepresented, $P = 0.10$ and $\Delta P = -0.330$. Intensive network-building is therefore not a pre-exit condition. Rather, it works as a relational reorganization state through which users remain within, or are redistributed across, the active participation space. This reconfiguration role also appears in patterns of re-entry: direct transitions from the Gap to High-intensity Networking are rare and underrepresented, $P = 0.02$ and $\Delta P = -0.013$, but $51.2\%$ of non-self transitions entering High-intensity Networking originate from the Gap. Intensive network-building can therefore also operate as an episodic mode of reactivation, without implying that inactivity generally leads preferentially to this state.
\noindent Low-intensity Networkers display a more boundary-adjacent pattern. In observed flows, the Gap is their largest destination, $P = 0.43$, indicating proximity to observable inactivity. However, this transition is underrepresented relative to the temporal-randomization null model, $\Delta P = -0.097$. By contrast, transitions toward Low-intensity Likers, $\Delta P = +0.041$, and Low-intensity Posters, $\Delta P = +0.016$, are overrepresented. Low-intensity networking is therefore close to the inactivity boundary in absolute terms, but it is not exit-attracted relative to the null model. Its distinctive temporal role is more local: it redirects users within the low-intensity active region rather than across the full participation space.
\noindent This contrast shows that network-building does not have a uniform temporal meaning. At high intensity, it functions as a broad reconfiguration hub across the active space and as an occasional reactivation state; at low intensity, it remains a more bounded form of reconfiguration, redirecting users locally within the low-intensity region while staying close to the inactivity boundary in observed flows.

\paragraph*{Local Oscillation}Low-intensity Likers and Low-intensity Posters form a bounded oscillatory regime. Both are self-persistent, with \(\Delta P = +0.129\) and \(\Delta P = +0.102\), respectively, and they show overrepresented reciprocal transitions: Low-intensity Likers to Low-intensity Posters, \(\Delta P = +0.013\), and Low-intensity Posters to Low-intensity Likers, \(\Delta P = +0.016\). The oscillation is asymmetric: movement from Low-intensity Posters to Low-intensity Likers, \(P = 0.13\), is stronger than the reverse, \(P = 0.08\), suggesting that low-intensity posting more often gives way to lightweight validation than the reverse.This regime is also close to inactivity. After self-transitions, the most likely destination from both states is the Gap, with \(P = 0.28\) for Low-intensity Likers and \(P = 0.27\) for Low-intensity Posters. At the same time, Low-intensity Likers receive incoming flows from High-intensity Likers, \(P = 0.14\), Low-intensity Networkers, \(P = 0.11\), and the Gap, \(P = 0.22\). The low-intensity liking/posting region therefore serves both as a local oscillation zone and as a landing area for users descending from higher intensity, exiting low-intensity networking, or returning from inactivity.

\paragraph*{Inertial Boundary}The Gap is strongly self-persistent, with \(\Delta P = +0.173\) and \(P = 0.77\). Transitions from active configurations into the Gap are underrepresented, and transitions from the Gap back into active configurations are also underrepresented. Inactivity therefore does not function as a basin into which active users disproportionately flow. Rather, it acts as an inertial boundary: once reached, it tends to persist and selectively limits re-entry.When re-entry occurs, it is concentrated in low-intensity configurations. The most probable destinations from the Gap are Low-intensity Likers and Low-intensity Networkers, both with \(P = 0.07\), while high-intensity configurations receive only negligible direct flows, \(P \leq 0.02\). The main exception is High-intensity Networking, for which \(40\%\) of incoming users come from the Gap. Thus, return from inactivity is usually low-investment, but it can also take the form of discontinuous reactivation through intensive network-building. Overall, the joint transition matrix shows that Bluesky participation is organized by differentiated dynamic regimes rather than by a linear movement from inactivity to activity. High-intensity liking and posting stabilize participation; high-intensity networking redirects users across the active space; low-intensity liking and posting form a bounded oscillatory zone; and the Gap acts as a persistent boundary that channels most re-entry through low-investment engagement, with the notable exception of intensive network-building.

\subsection*{Higher-order Profile Evolution}
\begin{figure*}
    \centering
    \includegraphics[width=0.7\linewidth]{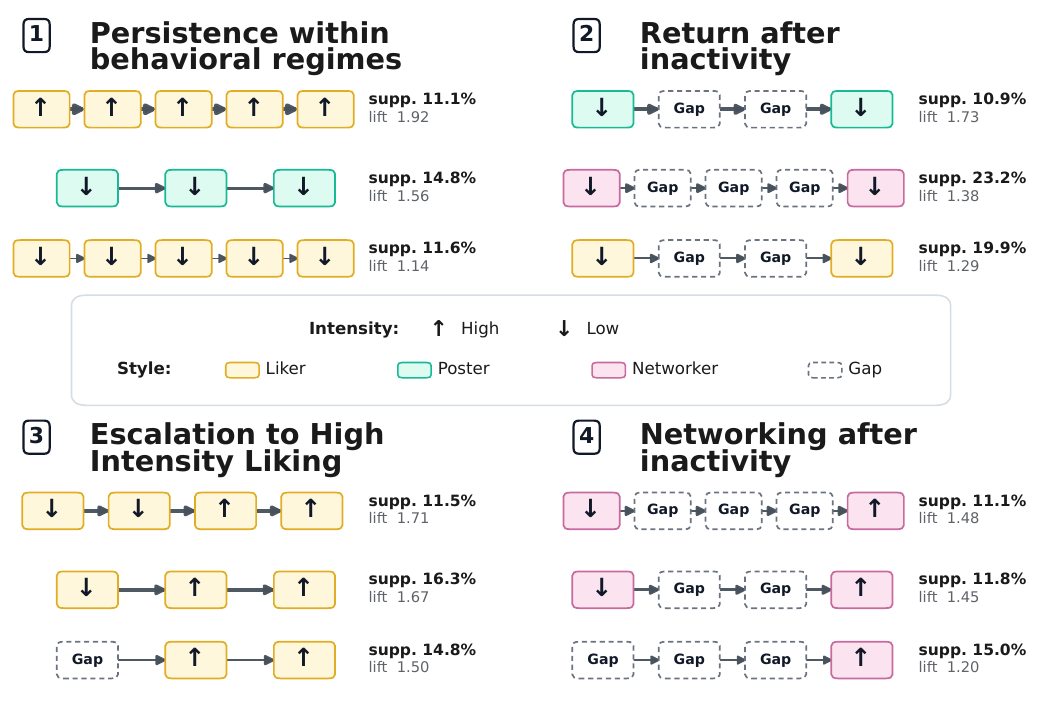}
    \caption{Frequent higher-order trajectory motifs identified with the PrefixSpan algorithm, together with representative examples, user support, and lift. We find that user trajectories can be summarized by four recurrent higher-order transition families.}
    \label{fig:patterns}
\end{figure*}

The sequential pattern analysis reveals that a small set of recurrent higher-order motifs structures user trajectories. 
As summarized in Fig.~\ref{fig:patterns}, these motifs can be grouped into four main families.
Further patterns are available in Table S1 in the SI.

\begin{resultbox}
\small
\textbf{Key results}: 
\begin{itemize}
    \item User trajectories are organized by recurrent higher-order motifs of persistence, interruption, return, and escalation.
    \item Gaps do not necessarily erase prior participation regimes: they can be followed by return to the same profile or by intensified re-entry.
    \item Low-intensity liking can precede durable high-intensity liking, while observed inactivity can precede intensified network-building.
\end{itemize}
\end{resultbox}

First, we observe persistence within behavioral regimes. 
This pattern is most pronounced for \emph{high-intensity Likers}, which exhibit the strongest multi-period self-reinforcement among the detected motifs. 
Persistence is also observed for \emph{low-intensity Posters} and \emph{low-intensity Likers}, although with lower lift values. 
This indicates that some behavioral profiles are not merely transient states, but tend to reproduce themselves over multiple observation periods.
Second, the results show return to the same regime after inactivity. 
The detected motifs indicate that Gaps often interrupt, rather than erase, prior behavioral regimes. 
This pattern is especially visible for low-intensity profiles, including \emph{low-intensity Posters}, \emph{low-intensity Networkers}, and \emph{low-intensity Likers}. 
Thus, inactivity should not be interpreted simply as noise or attrition: for a nontrivial share of users, it is followed by re-entry into the same behavioral profile.
Third, we identify movement into sustained \emph{high-intensity Liking}. 
These motifs show that transitions into high-intensity liking are frequently followed by persistence in the same regime. 
In particular, users moving from \emph{low-intensity Liking} into \emph{high-intensity Liking} tend to remain in the latter state for multiple periods. 
This suggests an asymmetric escalation pathway, whereby low-intensity liking can precede a more durable high-intensity engagement regime.
Finally, we identify a fourth family: movement into \emph{high-intensity Networking} after inactivity. 
This pattern links low-intensity networking and Gap periods to later high-intensity networking. 
It suggests that inactivity can also precede shifts toward more intensive network-building behavior, rather than only returns to prior low-intensity regimes.

Overall, the higher-order analysis complements the first-order transition results by showing that user trajectories are organized also by multi-period motifs of persistence, interruption, return, and escalation.

\section*{Discussion and Conclusions} 
This study started from a simple limitation of post-centered accounts of online participation: the traces through which platform activity is most visible are not necessarily the traces through which platform presence is most commonly sustained. Vocal production is highly concentrated, yet this concentration does not map onto a simple opposition between active contributors and absent or passive users. Many users participate through low-vocal actions that do not primarily add new content to the platform, but that nonetheless organize attention, visibility, relational exposure, and continuity of use.\\

\noindent Our results therefore suggest a shift from measuring participation as public expression to understanding it as platform presence and vocality shades: a temporally patterned position in an action space structured by platform affordances. In this space, intensity and style are analytically distinct. Users may be highly present without being highly vocal, and they may produce content without sustaining high-intensity engagement. The strongest association between high-intensity activity and liking shows that durable engagement is often organized around routinized validation rather than original content production. At the same time, posting-oriented users are relatively more common among low-intensity participants, suggesting that visibility and sustained engagement should not be conflated. Together, these findings reinforce the value of distinguishing participation style from participation intensity: how users participate and how much they participate capture distinct dimensions of platform presence that cannot be reduced to one another. Likes are low-cost actions, but they are not socially weak actions; on the contrary, they provide feedback, shape visibility, and stabilize everyday participation.\\

\noindent Network-building reveals a second form of low-vocal presence. Following is not merely a peripheral or preparatory action; it reorganizes the relational and informational environment in which future participation takes place. The distinction between high- and low-intensity Networkers is especially important. At high intensity, network-building functions as a reconfiguration hub that redirects users across the active space. At low intensity, it remains closer to inactivity, yet it still operates as a more bounded form of reorganization. This suggests that the same style of action can have different temporal roles depending on the intensity with which it is enacted.\\

\noindent The analysis of Gaps further shows that inactivity should not be treated as a simple null state. Observable inactivity is strongly persistent and selectively limits re-entry, but higher-order motifs show that it may also interrupt rather than erase prior regimes of participation. Users can return to the same low-intensity profile after a period of inactivity, and in some cases inactivity precedes intensive network-building. Platform participation is therefore not only a matter of how much users contribute, but of how they stabilize, suspend, resume, and transform their presence over time.\\

\noindent These findings contribute to the study of digital social life by challenging the residual treatment of silence. What is often classified as passive, silent, or missing may instead consist of differentiated low-vocality practices with distinct temporal roles.
For computational social science, this means that public expression should not be treated as a general proxy for participation. Models centered only on posts risk overrepresenting the most vocal users while underestimating the practices through which larger populations validate content, reorganize their social graph, and maintain intermittent forms of presence. For platform governance and design, the results suggest that metrics focused on content production alone miss important forms of engagement that sustain attention, visibility, and re-entry. More broadly, these findings highlight how the uneven observability of platform data can shape sociological inference. Because the most visible traces are disproportionately produced by vocal minorities, accounts of online participation built primarily around posting behavior risk overlooking the practices through which larger populations sustain platform life. Incorporating low-vocality forms of engagement can therefore help counter this visibility bias and provide a more representative understanding of how participation is organized and maintained.\\

\noindent Several limitations follow from the behavioral nature of the data. We observe platform traces, not motivations, interpretations, or unrecorded acts of reading. Low observable interaction should therefore not be equated directly with psychological passivity, disengagement, or silence. Future work should connect participation profiles to content domains, communities, and information diets, and should combine behavioral traces with surveys, interviews, or experiments capable of explaining why users remain low-vocal. Comparative analyses across platforms will also be necessary to determine which regimes are specific to Bluesky’s architecture and which reflect more general dynamics of online participation. More broadly, the results call for a theory of participation that treats low-vocality not as the absence of expression, but as one of the ordinary ways in which social platforms are inhabited.\\

\noindent These results have implications for the study of digital social life. They suggest that the analytical opposition between active users and passive users is too coarse to capture how platform environments are inhabited. What is often treated as silence or passivity may instead consist of differentiated, recurrent forms of participation. In this sense, the so-called silent majority is not simply silent: it is distributed across modes of low-vocal presence that validate, curate, interrupt, and reactivate platform life.

\section*{Materials and Methods}
\subsection*{Data Collection}
Data for our study are collected via the Bluesky API. 
Bluesky stores user data in Personal Data Servers (PDSes), which are publicly accessible repositories containing the full history of a user’s activity on the platform. 
Starting from a sample of approximately $5$ million users~\cite{failla2024m}, we query their PDSes to retrieve all available activity records up to January 2025. 
The dataset includes multiple action types performed by users, such as follows, likes, posts, reposts, and blocks, among others.
Please refer to Table~\ref{tab:actions} for a full description.

\begin{table}[htbp]
\centering
\small
\caption{Bluesky activity records}
\label{tab:actions}
\begin{tabular}{p{0.24\linewidth}p{0.48\linewidth}r}
\hline
\textbf{Action} & \textbf{Description} & \textbf{Count} \\
\hline
like & Likes on posts, feeds, etc. & 1,987,936,479 \\
post & Bluesky posts. Includes quotes. & 448,056,212 \\
follow & Follow relationships. & 419,344,962 \\
repost & Reposts of existing posts. & 234,617,365 \\
block & Account blocks. & 37,267,071 \\
list item & Addition of an account to a list. & 14,823,406 \\
list block & Blocks of entire lists. & 1,893,734 \\
threadgate & Thread reply controls. & 1,625,245 \\
postgate & Post interaction rules. & 517,144 \\
profile & Account profile declarations. & 380,564 \\
list & Account lists. & 354,506 \\
starter pack & Starter packs of actors and feeds. & 74,106 \\
feed generator & Feed generator metadata. & 61,668 \\
chat declaration & Chat account declarations. & 2 \\
\hline
\textbf{Total} & & \textbf{3,146,952,464} \\
\hline
\end{tabular}

\end{table}

\subsection*{User Session Estimation}
We approximate the time windows during which users are active on the platform by reconstructing user sessions from their activity logs. 
Following publicly reported statistics indicating that the average Bluesky session lasts approximately 10 minutes~\cite{explodingtopicsBlueskyUser}, we define sessions by grouping actions that occur within a $\pm 5$ minute window. 
Consecutive activity windows are merged when they overlap or are separated by short Gaps, yielding contiguous session intervals that approximate periods of continuous user engagement.
Our approach is fundamentally a stricter variant on a popular sessionization approach~\cite{halfaker2015user}
With this heuristic, we detect $\sim 74$M sessions across 4M users between January 2023 and January 2025 (average: 17, median: 8).
\subsection*{Mapping Participation Dimensions: Intensity Features} To characterize user behavior on the platform, we construct a two-dimensional representation of participation that separates \emph{intensity} (how much users engage) from \emph{style} (how users distribute their activity across actions). 
This distinction allows us to disentangle volumetric aspects of participation from compositional ones, preserving the interpretability of both dimensions while reducing the complexity of the behavioral space.
All features are derived from user-level activity aggregated over an observation period of one month. 
For each user \(u\) and monthly observation period \(t\), we compute both summary statistics capturing overall activity and disaggregated counts of different action types. 
These quantities are then used to construct two separate feature spaces corresponding to participation intensity and style.\\
The intensity dimension captures the degree of user presence and engagement on the platform. 
We operationalize intensity using five complementary features that reflect access frequency, activity volume, temporal investment, and within-session behavior.
Let \(S_{u,t}\) denote the total number of sessions, \(A_{u,t}\) the total number of actions, and \(D_{u,t}\) the total time spent on the platform (in minutes) by user \(u\) during period \(t\). We define the intensity feature vector as:
\begin{equation}
\mathbf{x}^{(\mathrm{int})}_{u,t}
=
\log\!\Big(
1 +
\big(
S_{u,t},\,
A_{u,t},\,
D_{u,t},\,
\frac{A_{u,t}}{S_{u,t}},\,
\frac{D_{u,t}}{S_{u,t}}
\big)
\Big)
\in \mathbb{R}^5,
\end{equation}
These features correspond to: (i) \emph{presence}, measured by the number of sessions; (ii) \emph{activity volume}, measured by the total number of actions; (iii) \emph{temporal immersion}, captured by total time spent; (iv) \emph{operational density}, defined as the average number of actions per session; and (v) \emph{average session duration}. 
All components are transformed via $\log(1+\cdot)$ to reduce skewness and limit the influence of extreme values.
Together, these dimensions distinguish, for example, users who access the platform frequently but perform few actions from those who concentrate high activity within fewer, more intensive sessions.

\subsection*{Mapping Participation Dimensions: Style Features}  While intensity captures the volume of activity, the style dimension characterizes how users allocate their actions across different types. 
To isolate compositional behavior from overall activity levels, we define style features in terms of relative action shares.
Let \(\mathcal{A}\) denote the set of action types (e.g., posting, liking, reposting, following). For each action \(a \in \mathcal{A}\), we define:
\begin{equation}
p_{u,t}(a) = \frac{c_{u,t}(a)}{A_{u,t}},
\qquad
A_{u,t} = \sum_{a \in \mathcal{A}} c_{u,t}(a),
\end{equation}
where \(c_{u,t}(a)\) is the number of actions of type \(a\) performed by user \(u\) in period \(t\). 
The quantity \(p_{u,t}(a)\) thus represents the proportion of activity devoted to action \(a\).
The resulting style feature vector is given by:
\begin{equation}
\mathbf{x}^{(\mathrm{sty})}_{u,t}
=
\bigl(p_{u,t}(a)\bigr)_{a \in \mathcal{A}}
\in \mathbb{R}^{|\mathcal{A}|}.
\end{equation}
By construction, this vector defines a normalized distribution over action types, allowing us to compare users independently of their total activity levels. 
This representation enables the identification of distinct behavioral patterns, such as users primarily engaged in content production, interaction, or passive consumption.
\subsection*{User Clustering} 
We employ unsupervised clustering to identify recurring patterns of user behavior in the feature spaces defined above. 
Clustering enables the detection of latent structures in the data without relying on predefined labels, grouping observations based on their similarity in the feature space.
We adopt the \textit{K-means} algorithm~\cite{macqueen1967multivariate} due to its scalability and interpretability. 
K-means is a partition-based clustering method that assigns each observation to one of \(K\) clusters by minimizing within-cluster variance. 
Formally, given a set of observations \(\{\mathbf{x}_n\}_{n=1}^{N}\), the algorithm seeks a partition \(\{C_k\}_{k=1}^{K}\) that minimizes the within-cluster sum of squares:
\begin{equation}
\sum_{k=1}^{K} \sum_{\mathbf{x}_i \in C_k} 
\|\mathbf{x}_i - \boldsymbol{\mu}_k\|^2,
\end{equation}
where \(C_k\) denotes the set of points assigned to cluster \(k\) and \(\boldsymbol{\mu}_k\) is the corresponding centroid. 
The algorithm proceeds iteratively by alternating between assigning each observation to the nearest centroid (using Euclidean distance) and updating centroids as the mean of assigned points, until convergence.
Notably, clustering is performed on user--time observations, where each pair \((u,t)\) represents a user in a given observation period. 
This formulation allows the same user to occupy different positions in the feature space over time and to transition across clusters, enabling the analysis of behavioral dynamics. 
The procedure is applied separately to the two feature spaces introduced above---intensity and style---so as to avoid conflating volumetric differences (how much users engage) with compositional ones (how users engage).
The number of clusters \(K\) is selected using the \textit{silhouette score}, an internal validation metric that jointly evaluates cluster cohesion and separation~\cite{rousseeuw1987silhouettes}. 
For each observation, the silhouette coefficient compares the average distance to points within the same cluster to the minimum average distance to points in other clusters. The resulting score lies in the interval \([-1,1]\), where higher values indicate better-defined clusters. 
Due to computational constraints, silhouette scores are computed on a random subsample of the data (X\%), while the final model is estimated on the full dataset. 
The optimal \(K\) is chosen as the value that maximizes the average silhouette score.
\subsection*{Identifying Statistically Significant Regime Transitions} 
We model the temporal evolution of user profiles as a discrete-time Markov process, where each cluster is a state and each user trajectory is a sequence of states over time. 
Let
\[
\mathbf{s}_u=(s_{u,1},\ldots,s_{u,T_u}),
\qquad
s_{u,t}\in\{1,\ldots,K,g\},
\]
denote the state sequence of user \(u\), where \(g\) represents inactivity or missing observations within the user’s active window.
From these sequences, we estimate the empirical transition probability from state \(i\) to state \(j\) as
\[
P_{ij}
=
\mathbb{P}(s_{t+1}=j\mid s_t=i)
=
\frac{N_{ij}}{\sum_{\ell}N_{i\ell}},
\]
where \(N_{ij}\) is the number of observed transitions from \(i\) to \(j\).
To assess statistical significance, we compare the observed transition matrix to a within-user permutation null model. 
For each permutation \(b=1,\ldots,B\), we randomly shuffle the order of states within each user trajectory,
\[
\mathbf{s}^{(b)}_u=\pi^{(b)}_u(\mathbf{s}_u),
\]
thereby preserving each user’s marginal state composition while removing temporal dependence. 
We then recompute the transition probabilities \(P^{(b)}_{ij}\), producing a null distribution for each transition \(i\to j\).
The deviation of the observed transition probability from the null model is summarized by
\[
z_{ij}
=
\frac{P_{ij}-\mu^{0}_{ij}}{\sigma^{0}_{ij}},
\]
where
\[
\mu^{0}_{ij}
=
\frac{1}{B}\sum_{b=1}^{B}P^{(b)}_{ij},
\qquad
\sigma^{0}_{ij}
=
\sqrt{
\frac{1}{B-1}
\sum_{b=1}^{B}
\left(P^{(b)}_{ij}-\mu^{0}_{ij}\right)^2
}.
\]
Empirical two-sided \(p\)-values are computed as
\[
p_{ij}
=
\frac{
1+
\sum_{b=1}^{B}
\mathbb{I}\!\left(
\left|P^{(b)}_{ij}-\mu^{0}_{ij}\right|
\ge
\left|P_{ij}-\mu^{0}_{ij}\right|
\right)
}{
B+1
}.
\]
This procedure identifies transitions that occur more or less frequently than expected under temporal independence. 
In the experimental campaign, we set \(B=100\).

\subsection*{Higher-order Profile Evolution} 
We also analyze higher-order sequential structure that unfold across multiple observation periods. 
We represent each user \(u\) as an ordered sequence of profile states,
\[
\mathbf{s}_u=(s_{u,1},\ldots,s_{u,T_u}),
\qquad
s_{u,t}\in\{1,\ldots,K,g\},
\]
where \(K\) denotes the number of behavioral clusters and \(g\) is treated as an explicit state representing inactivity or missing observations within the user’s active window. 
Thus, Gaps are not discarded but are included in the same state space as the behavioral clusters, allowing the analysis to capture trajectories that pass through periods of inactivity.

To identify recurrent multi-step trajectories, we apply sequential pattern mining to the collection of user sequences \(\{\mathbf{s}_u\}_{u=1}^{U}\). 
Specifically, we use the \textit{PrefixSpan} algorithm~\cite{pei2004mining}, a pattern-growth method that discovers frequent subsequences without explicitly generating candidate sequences.
Let \(\boldsymbol{\alpha}=(\alpha_1,\ldots,\alpha_m)\) denote a candidate sequential pattern of length \(m\). 
We say that \(\boldsymbol{\alpha}\) is contained in a user trajectory \(\mathbf{s}_u\) if there exist indices $1 \leq t_1 < t_2 < \cdots < t_m \leq T_u$
such that $s_{u,t_r}=\alpha_r \qquad \text{for all } r=1,\ldots,m$.
The support of pattern \(\boldsymbol{\alpha}\) is then defined as
\[
\mathrm{supp}(\boldsymbol{\alpha})
=
\sum_{u=1}^{U}
\mathbb{I}\!\left(\boldsymbol{\alpha}\preceq \mathbf{s}_u\right),
\]
where \(\boldsymbol{\alpha}\preceq \mathbf{s}_u\) indicates that \(\boldsymbol{\alpha}\) occurs as an ordered subsequence of \(\mathbf{s}_u\). 
The corresponding relative support is given by
\[
\mathrm{rsupp}(\boldsymbol{\alpha})
=
\frac{\mathrm{supp}(\boldsymbol{\alpha})}{U}.
\]
We retain frequent patterns with length \(3 \leq m \leq 5\), thereby focusing on higher-order trajectories that extend beyond single-step transitions while limiting very long and sparse sequences.
To distinguish common trajectories from unusually strong directional associations, we further compute rule-level metrics for each discovered pattern. 
For every split of a pattern \(\boldsymbol{\alpha}\) into an antecedent and consequent,
\[
\boldsymbol{\alpha}
=
(\boldsymbol{\alpha}^{-},\boldsymbol{\alpha}^{+}),
\]
confidence is defined as
\[
\mathrm{conf}
\left(
\boldsymbol{\alpha}^{-}\rightarrow \boldsymbol{\alpha}^{+}
\right)
=
\frac{
\mathrm{supp}(\boldsymbol{\alpha})
}{
\mathrm{supp}(\boldsymbol{\alpha}^{-})
},
\]
and lift as
\[
\mathrm{lift}
\left(
\boldsymbol{\alpha}^{-}\rightarrow \boldsymbol{\alpha}^{+}
\right)
=
\frac{
\mathrm{conf}
\left(
\boldsymbol{\alpha}^{-}\rightarrow \boldsymbol{\alpha}^{+}
\right)
}{
\mathrm{rsupp}(\boldsymbol{\alpha}^{+})
}.
\]
Confidence measures how often the consequent is observed among users who exhibit the antecedent, whereas lift compares this probability to the baseline prevalence of the consequent in the population. 
Values of lift greater than one indicate higher-order profile evolutions that occur more frequently than expected from marginal subsequence frequencies alone. 
Because \(g\) is included as a valid state, the resulting patterns can explicitly capture trajectories involving inactivity, such as transitions into, through, or out of Gaps.

\section*{Use of Generative Artificial Intelligence} The authors acknowledge the use of GPT5.5 by OpenAI for revising the article text. All generated text was reviewed by all authors. 
The authors also acknowledge the use of said model for generating boilerplate code. 
All generated code was reviewed by at least two authors.
The authors take full responsibility for the content of this article and its findings.

\section*{Data Availability}
We publicly release pseudo-anonymized data in the associated Zenodo repository~\cite{datasetLowVocality}. Code will be shared in the repository upon acceptance.

\bibliographystyle{unsrtnat}
\bibliography{references}

\clearpage
\onecolumn
\setcounter{figure}{0}
\setcounter{table}{0}
\renewcommand{\thefigure}{S\arabic{figure}}
\renewcommand{\thetable}{S\arabic{table}}
\section*{Supporting Information}
\section*{Characterization of User Sessions and Activity}

\begin{figure}[!htbp]
    \centering
    \includegraphics[width=\linewidth]{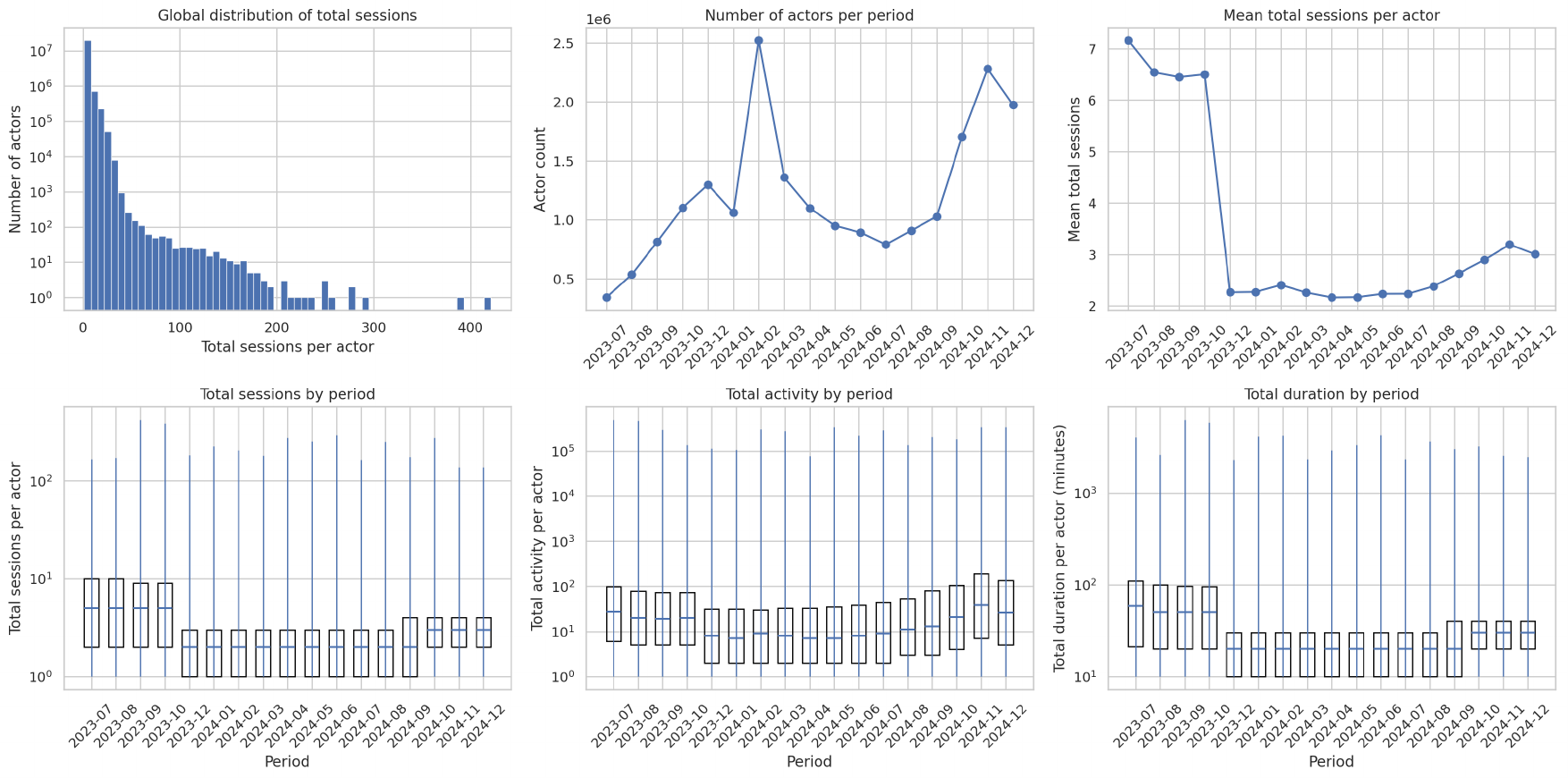}
    \caption{Top: distribution of sessions per author (left); monthly user trend (middle); average user session (right). Bottom: monthly distribution of user-wise sessions (left); monthly distribution of user-wise activities (middle); monthly distribution of user-wise total duration (right).}
    \label{fig:poster_stats}
\end{figure}

\noindent Before reconstructing behavioral profiles, we inspected the marginal and temporal distributions of the main actor-level variables used to describe platform presence: the number of reconstructed sessions, the total number of observed actions, and the total duration of reconstructed sessions in minutes. Fig.~\ref{fig:poster_stats} summarizes these variables over the observation window. November 2023 is not treated as a period of inactivity, but as an absent month in the temporal series, consistently with the data exclusion described in the main text.\\
It should be noted that due to a failure in our computational infrastructure, all and only observations from November 2023 were discarded. 
However, given the limited size of the affected sample (4.5\% of the session records), we estimate this exclusion to have no significant impact on the results.
In temporal settings, November 2023 is treated as absent from the observation period rather than as a period of user inactivity.

\noindent The global distribution of sessions is strongly right-skewed. Most actors are observed with only a small number of sessions, whereas a small minority reach substantially higher values. This long-tailed structure is consistent with the concentration of participation observed in the main analysis and motivates the use of log-scaled representations for the volumetric distributions.\\

\noindent The monthly panels show that changes in per-actor averages cannot be interpreted independently of changes in the size of the observed active population. The number of active actors increases markedly between the second half of 2023 and the beginning of 2024, followed by a contraction and then by a renewed increase in the second half of 2024. Over the same period, the mean number of sessions per actor declines sharply after the initial phase and remains comparatively low throughout the first half of 2024. This inverse movement suggests a compositional effect: as the observable active population expands, the entry of users with lower or more intermittent activity reduces aggregate per-actor averages, without necessarily implying a uniform decline in the activity of persistent users.\\

\noindent Similar dynamics are visible for total activity and total duration. Both variables display highly asymmetric monthly distributions, with most actors concentrated at low values and long upper tails corresponding to a small number of highly active users. After the higher-activity phase observed in 2023, the distributions shift downward during early and mid-2024 and then increase again toward the end of the observation window. The late-2024 increase is especially visible for total activity and total duration, suggesting a partial re-consolidation of more intensive usage patterns rather than a simple return to the initial configuration.\\

\noindent The lower bound visible in duration around the ten-minute scale, should be interpreted in relation to the session-reconstruction procedure. Since sessions are approximated by grouping timestamped actions into windows based on a ten-minute reference duration, the absence of very short reconstructed durations primarily reflects the operational definition of sessions rather than an independent behavioral threshold.\\

\noindent Overall, Fig.~\ref{fig:poster_stats} indicates that raw activity volume is not sufficient to characterize participation. The descriptive variables are highly skewed, and their monthly averages are strongly affected by changes in the size and composition of the active population. This motivates the longitudinal user-level approach adopted in the main analysis, where usage intensity is separated from participation style in order to distinguish how much users engage from how they distribute their activity across platform practices.

\section*{Temporal Dynamics of User Profiles}
In the following, we report a longitudinal volumetric description of the data with respect to user profiles.
\subsection*{Intensity}
\begin{figure}[!htbp]
\centering
\includegraphics[width=\linewidth]{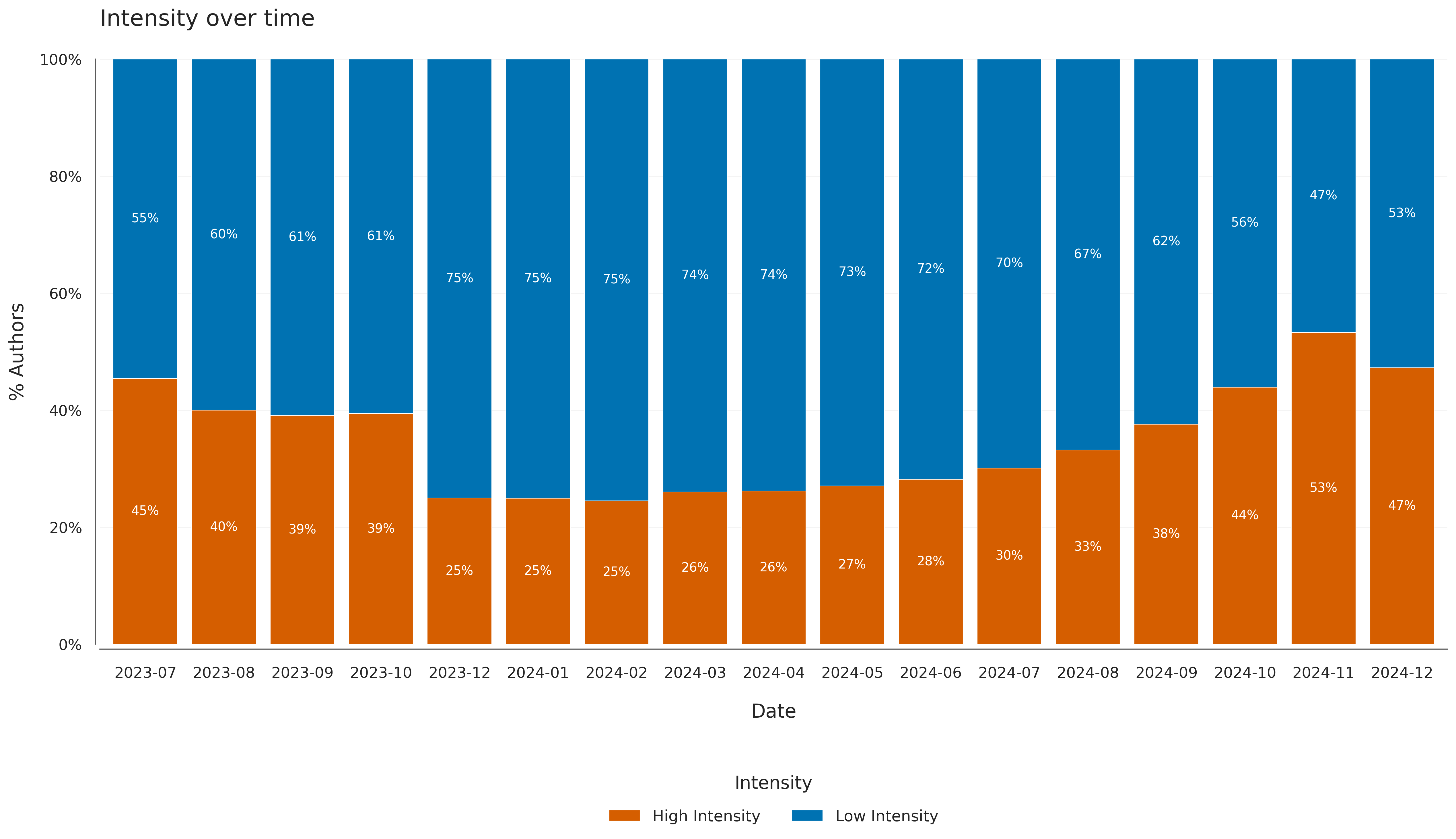}
\caption{Monthly share of users per intensity profile.}
\label{fig:log_cluster_volumes_over_time_epjds_stacked_bar}
\end{figure}

The analysis of the temporal evolution of usage-intensity cluster sizes in Fig.~\ref{fig:log_cluster_volumes_over_time_epjds_stacked_bar} reveals clearly distinct dynamics between the two groups, confirming that usage intensity is not a static property of the population, but rather varies over time. Over the period considered, the low-intensity cluster is largely dominant for most of the observation window, reaching a peak between late 2023 and early 2024, when it consistently exceeds 70\% of active users. During this phase, the platform appears to be characterized by a prevalence of low-operational-density forms of use, suggesting growth driven primarily by limited consumption and interaction practices.\\

\noindent From mid-2024 onward, however, a progressive reversal of this trend can be observed. The share of the low-intensity cluster gradually decreases, while that of the high-intensity cluster increases correspondingly, leading to a substantial convergence between the two groups during autumn 2024. This rebalancing culminates in November 2024, when the high-intensity cluster temporarily becomes predominant, exceeding 50\% of active users.\\

\noindent In the final observation period, the distribution appears to move toward a more balanced configuration, although with some fluctuations between the two clusters. This pattern may suggest a maturation process within the ecosystem, whereby a growing share of the user base develops more intensive and routinized usage patterns, reducing the relative weight of low-density forms of participation. Overall, the temporal dynamics of cluster volumes confirms that usage intensity is sensitive to structural changes in the platform and represents a relevant indicator of the consolidation or reorientation of participatory practices.

\subsection*{Style}
\begin{figure}[!htbp]
\centering
\includegraphics[width=\linewidth]{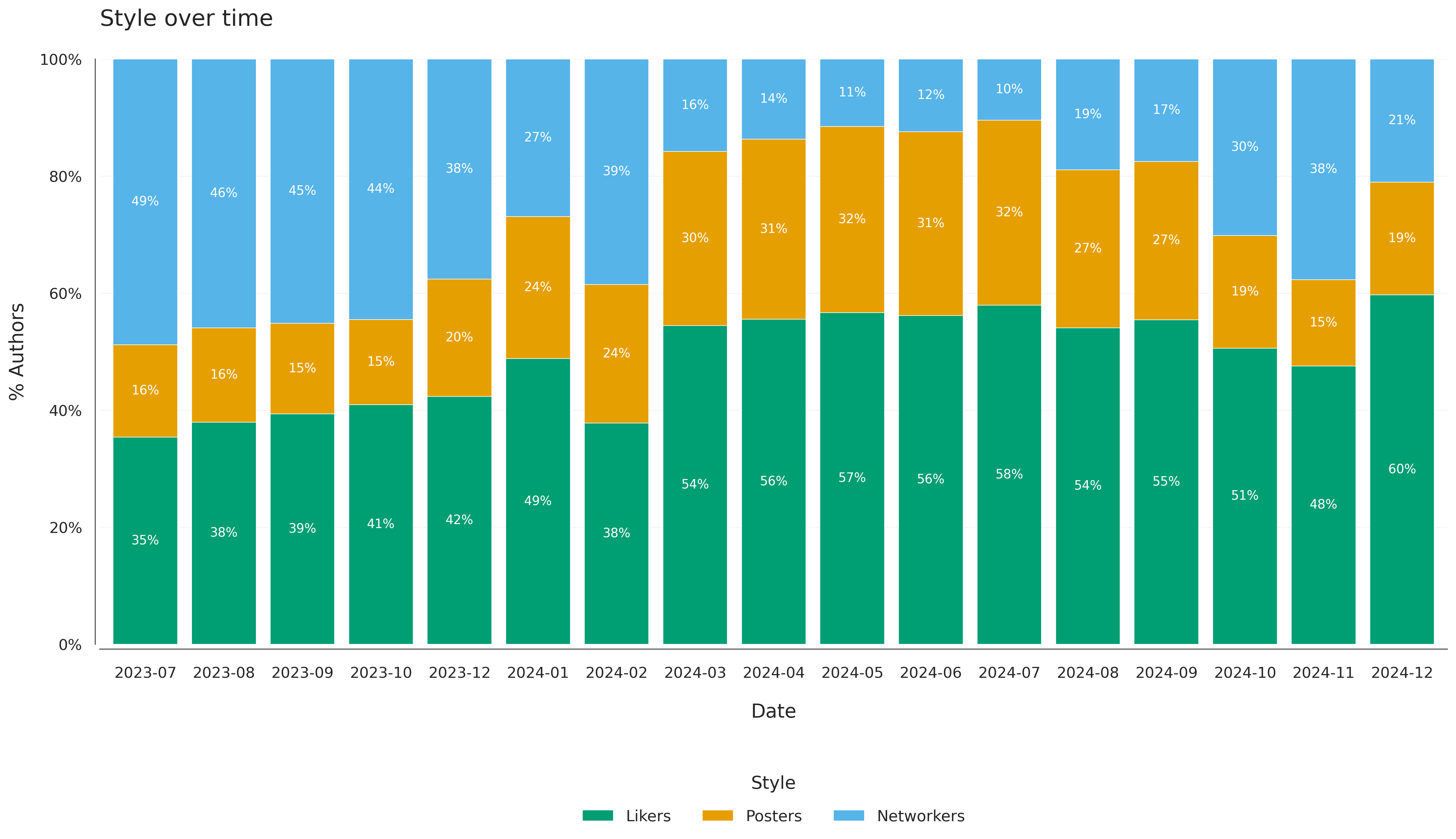}
\caption{Monthly share of users per style profile.}
\label{fig:share_cluster_volumes_over_time_epjds_stacked_bar.png}
\end{figure}

The analysis of the temporal volumes of the participation-style clusters reveals a complex dynamic, marked by a progressive reorganization of the behavioral composition of the active population. In the initial observation period, the distribution appears relatively balanced between \textit{Likers} and \textit{Networkers}, with the latter representing the largest share of active users, while the \textit{Posters} cluster occupies a clearly minority but relatively stable position. During this period, the \textit{Likers} cluster shows a gradual increase, rising from about 35\% to over 40\% of active users, accompanied by a progressive decline in the share of \textit{Networkers}. This dynamic suggests an initial process of platform adjustment after the early adoption phase, in which part of the user base appears to shift from network-building practices toward more routinized forms of participation, primarily centered on lightweight content validation.\\

\noindent However, the beginning of 2024 is marked by a light volatility. In particular, after a temporary rebound in February, the \textit{Networkers} cluster sharply contracts from March 2024 onward, falling below 20\% of active users. At the same time, the \textit{Likers} cluster records a rapid increase, exceeding the 50\% threshold for most of the following months. In this phase, the \textit{Likers} style emerges as the dominant mode of participation, absorbing a growing share of the active population.\\

\noindent The \textit{Posters} cluster, by contrast, follows a more contained but still relevant dynamic. After a gradual increase between late 2023 and the first half of 2024, reaching around 30\% of active users, its incidence tends to decline in the second half of the year, while remaining a relevant component of the active population. This pattern suggests that content-oriented participation remains structurally minority and subject to fluctuations, rather than following a cumulative growth trajectory over time.\\

\noindent In the final observation period, the distribution appears more unstable again, with a temporary recovery of \textit{Networkers} between October and November and a decline in \textit{Likers} over the same interval, followed by a renewed increase in \textit{Likers} in the last observed month. Overall, the analysis of temporal volumes indicates that participation style is not a fixed characteristic of users, but a dynamic dimension: practices oriented toward network building tend to decline over the course of 2024 in favor of more routinized modes of consumption and validation, while content production remains a minority but still relevant practice.

\subsection*{Configurations}

\begin{figure}[!htbp]
\centering
\includegraphics[width=\linewidth]{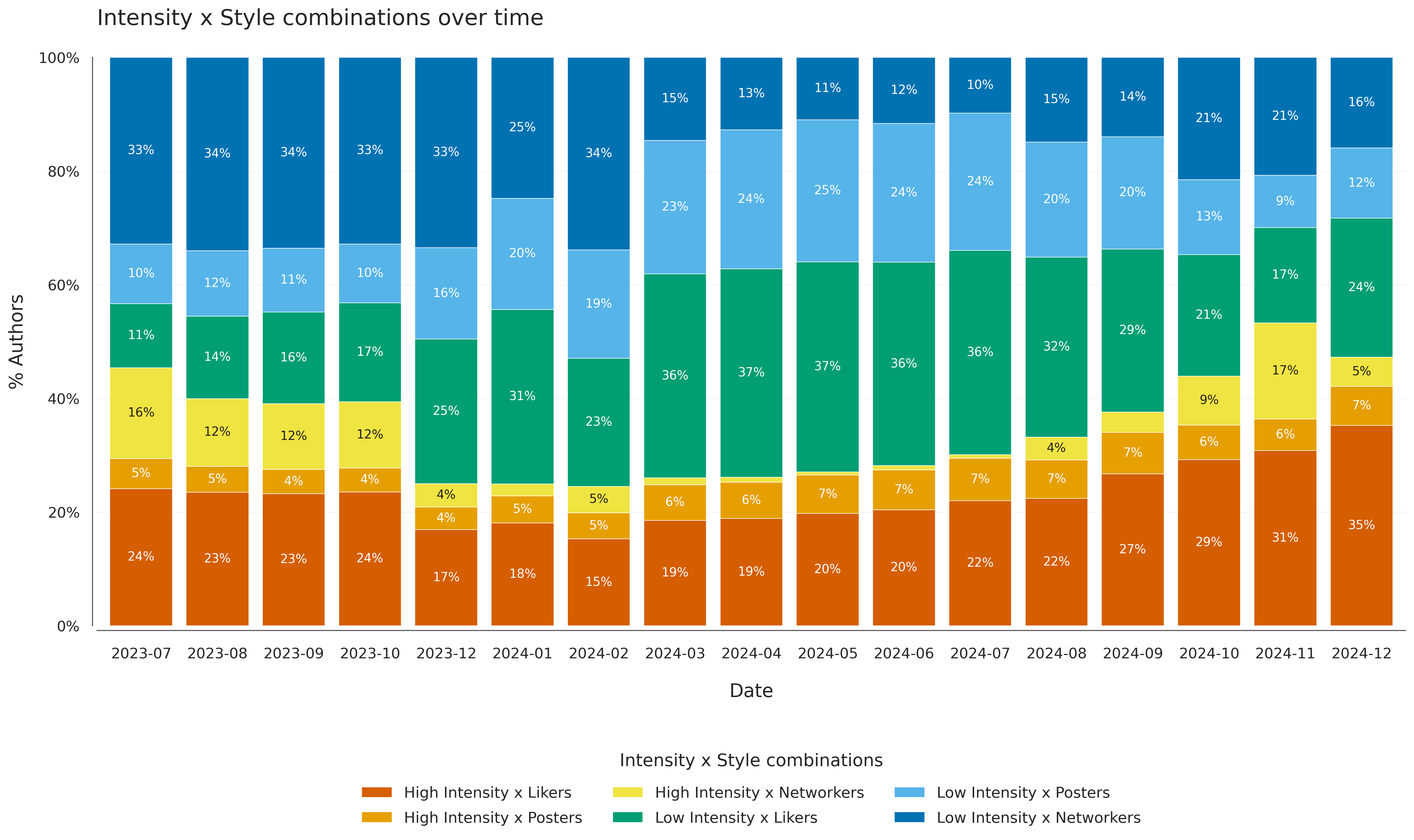}
\caption{Monthly share of users per style-intensity profile.}
\label{fig:log_share_combination_epjds_stacked_bar}
\end{figure}

The monthly evolution of intensity--style combinations in Fig.~\ref{fig:log_share_combination_epjds_stacked_bar} makes it possible to observe the compositional dynamics of the space over the period analyzed. This representation is based on the same set of actor--month observations, but based on the intersection of the two clustering dimensions month by month, thus making visible changes in the relative distribution of positions over time. In the initial phase, corresponding to the second half of 2023, the \textit{Low Intensity} $\times$ \textit{Networkers} configuration represents the largest position, accounting for approximately 33--34\% of active users. This indicates that, during the early consolidation phase of the platform, participation is largely characterized by network-building practices associated with low-intensity forms of use.\\

\noindent From the first months of 2024 onward, the relative weight of this configuration progressively decreases, accompanied by the growth of the \textit{Low Intensity} $\times$ \textit{Likers} combination, which exceeds 35\% between March and July 2024. In this phase, the space appears to reorganize around forms of participation more oriented toward lightweight content validation, while still predominantly maintaining a low level of intensity. During the second half of 2024, a further compositional shift can be observed: the \textit{High Intensity} $\times$ \textit{Likers} configuration grows progressively, reaching about 35\% in the last observed month and becoming the dominant position in the overall distribution. This evolution points to a growing concentration of participation around configurations characterized by higher operational investment and routinized validation.\\

\noindent Configurations associated with \textit{Posters}, by contrast, display more variable patterns. The \textit{Low Intensity} $\times$ \textit{Posters} combination increases during the first half of 2024, reaching values around 24--25\%, before declining in the second half of the year. The \textit{High Intensity} $\times$ \textit{Posters} configuration remains minority throughout the entire period, although it shows a slight increase in the final phase. Configurations associated with \textit{Networkers} also tend to decline overall compared with the initial phase, although the \textit{High Intensity} $\times$ \textit{Networkers} component exhibits some fluctuations and a temporary recovery in the second half of 2024.\\

\noindent Overall, these variations describe a progressive transformation in the composition of the space of practices, shaped by changes in the prevailing forms of participation over time. While the set of possible configurations remains unchanged, their relative distribution shifts, producing a redefinition of the internal hierarchies among network-building, lightweight validation, and content-production practices. The space of practices does not change in its formal structure, but rather in the concrete configuration of its participatory equilibrium.

\section*{Usage Intensity Trajectories}
\begin{figure}[t]
\centering
\includegraphics[width=\linewidth]{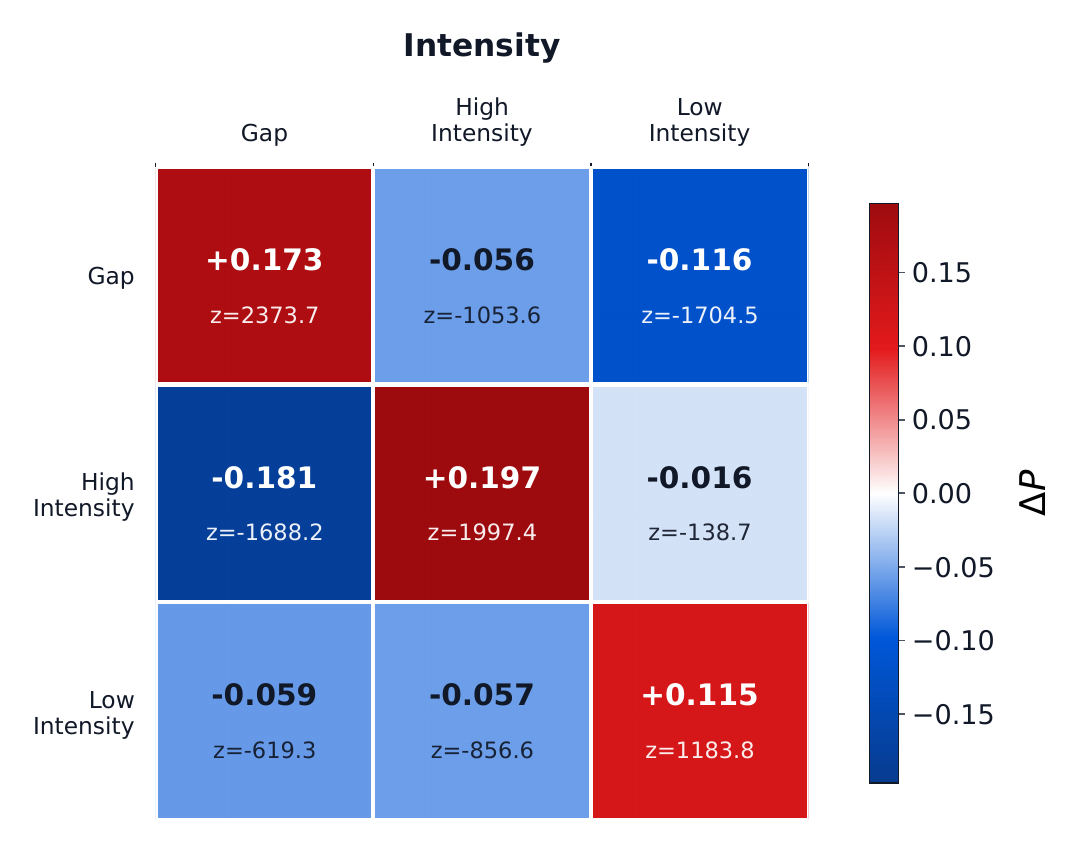}
\caption{Statistically significant first-order transitions between intensity clusters, expressed as deviations from the temporal-randomization null model. Values indicate $\Delta P = P_{\mathrm{obs}} - P_{\mathrm{null}}$. Red and blue cells indicate transitions that occur more and less frequently than expected, respectively.}
\label{fig:intensity-transitions}
\end{figure}

The intensity transition matrix in Fig.~\ref{fig:intensity-transitions} shows strong temporal inertia. All transitions differ significantly from the null model: self-transitions are overrepresented, while transitions between distinct states are underrepresented. Users thus remain in the same intensity state more often than expected under temporal independence, whereas shifts among inactivity, high-intensity, and low-intensity occur less often than chance would predict.
Both active intensity states therefore behave as self-reproducing positions in the platform’s participation space. High-intensity shows the strongest positive deviation among active states ($\Delta P=+0.197$), suggesting that dense operational presence tends to stabilize over time. Low-intensity also exhibits significant persistence ($\Delta P=+0.115$), indicating a less dense but still temporally organized mode of participation.
Absence likewise shows marked self-persistence ($\Delta P=+0.173$): discontinuity tends to reproduce itself, while returns to active states are less frequent than expected. Active states, however, are not especially prone to transition into inactivity. Transitions from both high and low-intensity to the Gap are also less frequent than expected, with the strongest negative deviation from high-intensity to the Gap ($\Delta P=-0.181$). Inactivity, therefore, functions as a persistent boundary condition: once reached, it tends to endure, while re-entry into observable participation remains selective.

\section*{Participation Style Trajectories}
\begin{figure}[t]
\centering
\includegraphics[width=\linewidth]{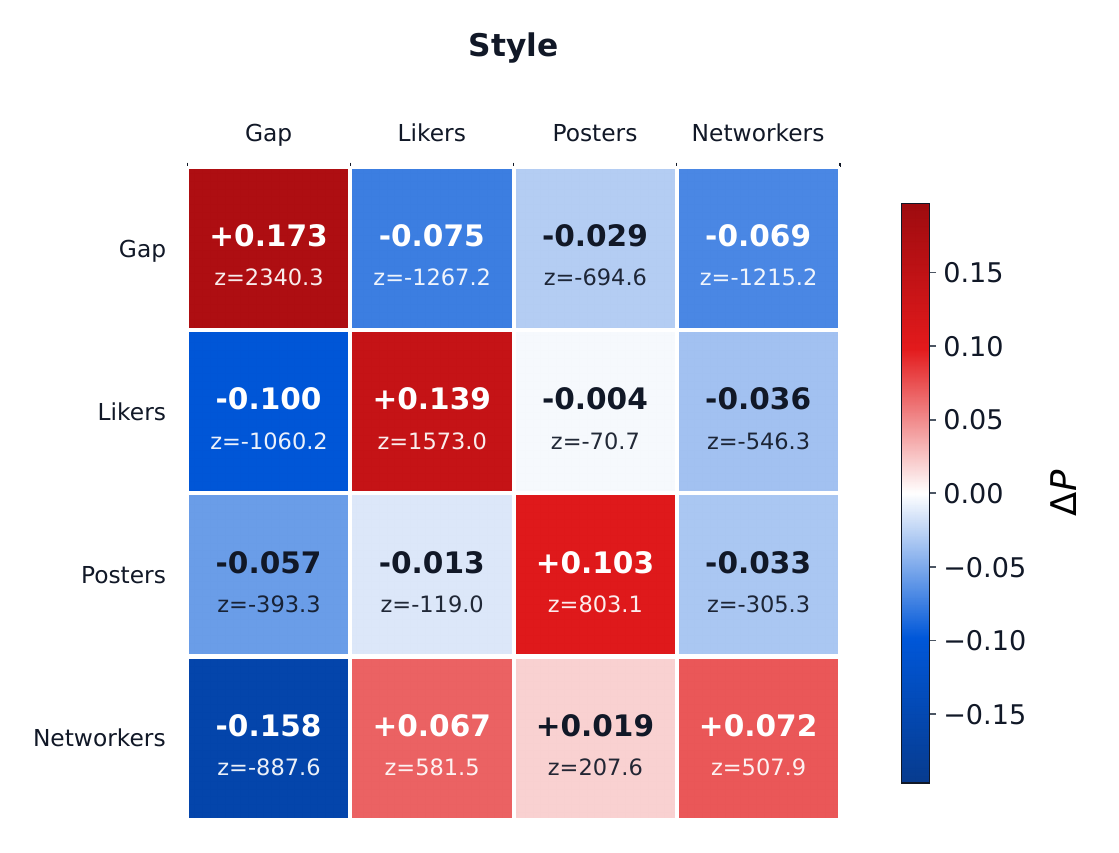}
\caption{Statistically significant first-order transitions between style clusters, expressed as deviations from the temporal-randomization null model. Values indicate $\Delta P = P_{\mathrm{obs}} - P_{\mathrm{null}}$. Red and blue cells indicate transitions that occur more and less frequently than expected, respectively.}
\label{fig:style-transitions}
\end{figure}

The participation style transition matrix in Fig.~\ref{fig:style-transitions} shows that the qualitative form of action is also temporally organized. Liking-oriented, posting-oriented, and network-building practices all exhibit significant self-persistence relative to the null model, indicating that styles are recurrent repertoires through which users inhabit the platform. However, these repertoires play different dynamic roles.
\\
Likers and Posters appear as comparatively closed forms of participation. Both states display overrepresented self-transitions, with a stronger deviation for Likers ($\Delta P=+0.139$) than for Posters ($\Delta P=+0.103$), while transitions from either state toward alternative styles are systematically underrepresented. Lightweight validation and content production therefore operate as stable repertoires: once users organize their activity around endorsement or posting, they are more likely to reproduce that mode of action than to shift toward a different participatory orientation.
\\
Networkers follow a different pattern. Although network-building also persists more than expected ($\Delta P=+0.072$), it is the only active style whose outgoing transitions toward other active styles are overrepresented. Transitions from Networkers to Likers ($\Delta P=+0.067$) and Posters ($\Delta P=+0.019$) occur more frequently than under the null model, while transitions to the Gap are strongly underrepresented ($\Delta P=-0.158$). Thus, network-building is not strictly a pre-exit condition. Despite being dominated by user-to-user relational actions rather than direct content engagement, it redirects users toward more content-oriented repertoires of validation or production.
\\
The style dimension therefore reveals a differentiated temporal morphology. Liking and posting provide stable modes of participation, whereas networking operates as a transformative position within the active space. The construction of the social graph is not a pathway to disengagement, but a mechanism through which users reorient participation inside the platform.

\begin{figure}[t]
\centering
\includegraphics[width=\linewidth]{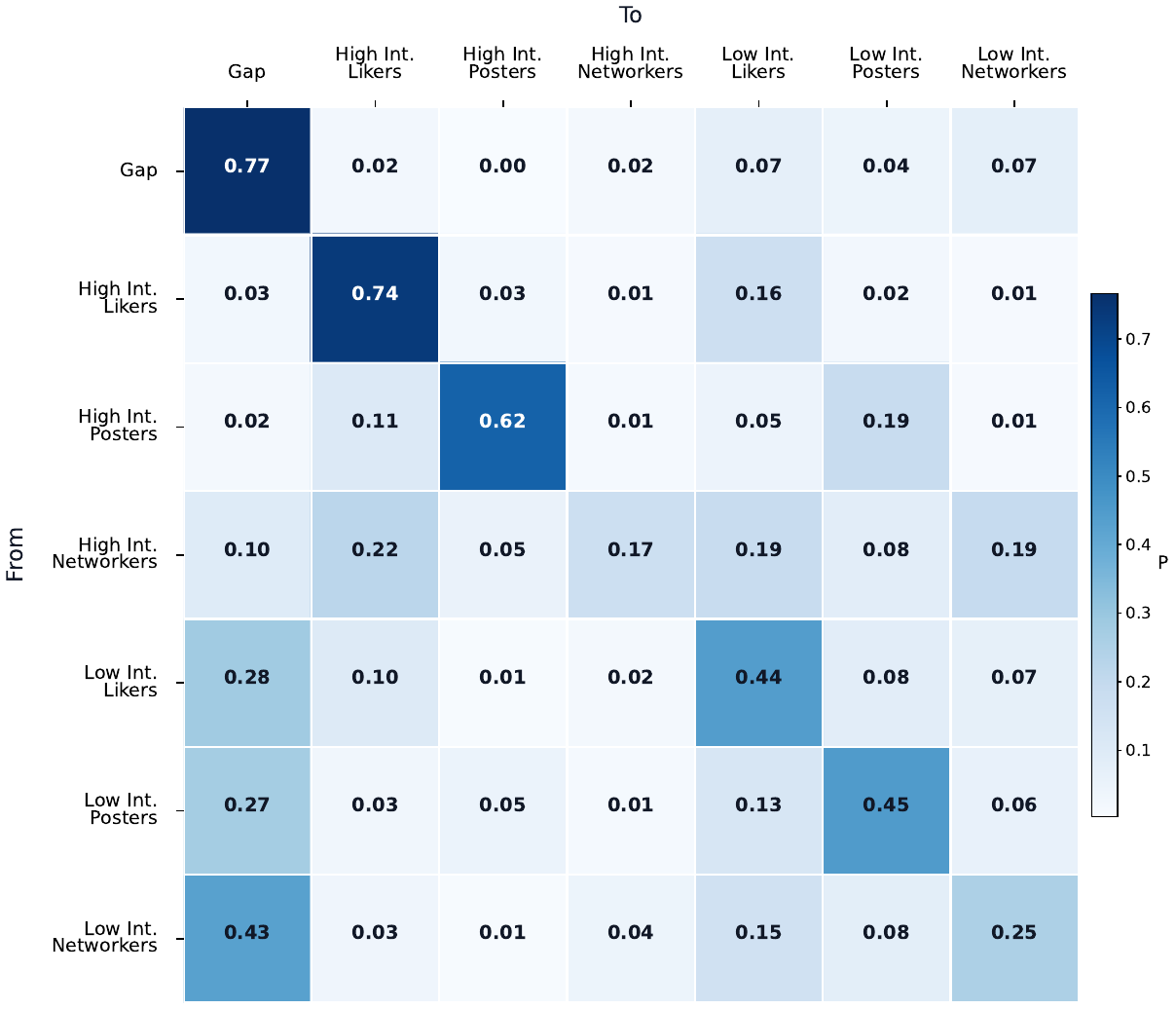}
\caption{Row-normalized transition probabilities across user profiles. Values are normalized within each source profile, so each row sums to one and each cell indicates the probability of transitioning from the corresponding source profile to the target profile.}
\label{fig:rownorm}
\end{figure}

\begin{figure}[t]
\centering
\includegraphics[width=\linewidth]{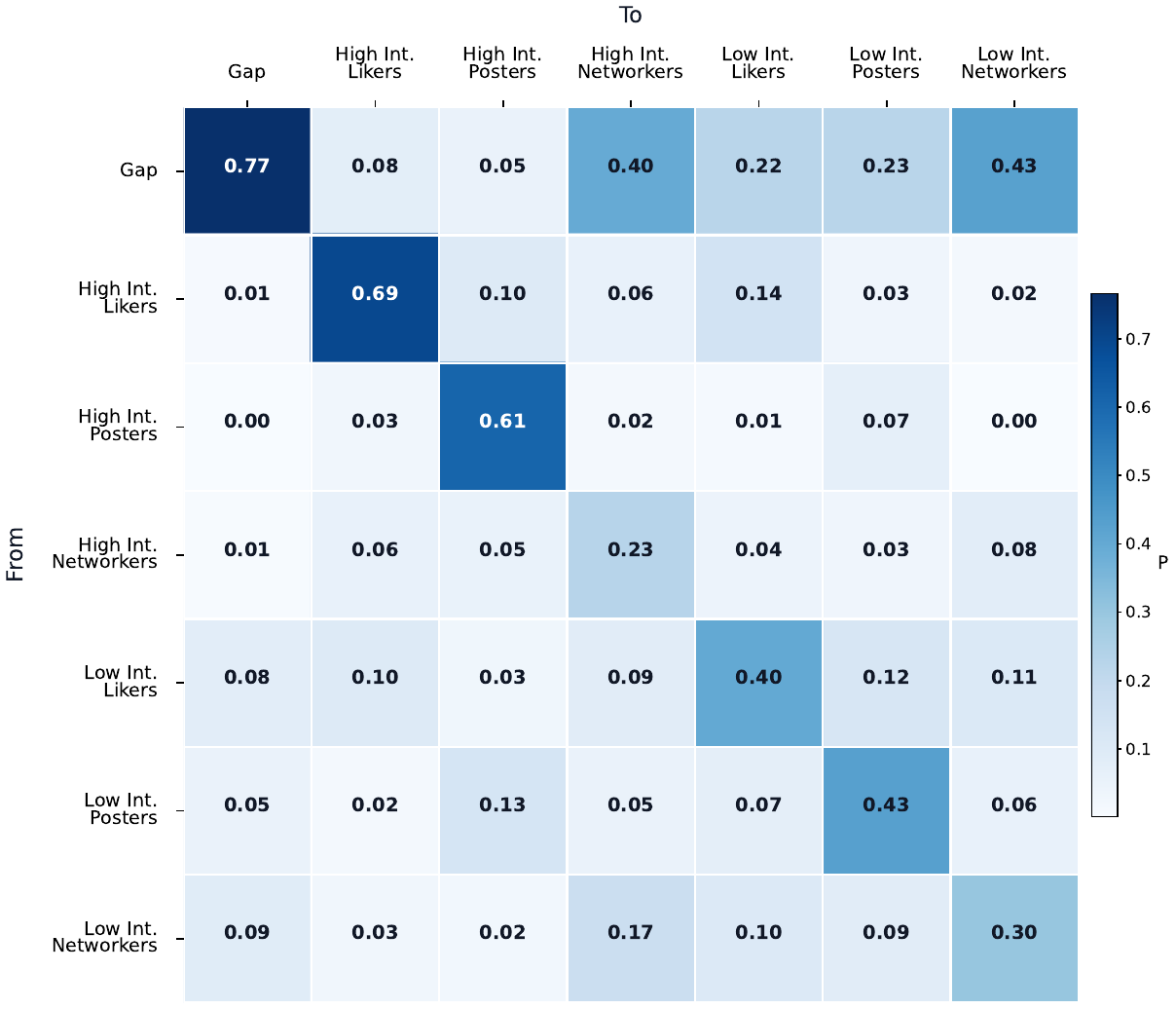}
\caption{Column-normalized transition probabilities across user profiles. Values are normalized within each target profile, so each column sums to one and each cell indicates the relative contribution of the corresponding source profile to the target profile.}
\label{fig:colnorm}
\end{figure}
\begin{table}[]
\centering
\begin{tabular}{rlrr}
\toprule
Rank & Pattern & Support (\%) & Lift \\
\midrule
1 & $\underline{\mathrm{L}^{\uparrow} \to \mathrm{L}^{\uparrow} \to \mathrm{L}^{\uparrow} \to \mathrm{L}^{\uparrow} \to \mathrm{L}^{\uparrow}}$ & 11.1 & 1.92 \\
2 & $\underline{\mathrm{P}^{\downarrow} \to \mathrm{Gap} \to \mathrm{Gap} \to \mathrm{P}^{\downarrow}}$ & 10.9 & 1.73 \\
3 & $\mathrm{L}^{\uparrow} \to \mathrm{L}^{\uparrow} \to \mathrm{L}^{\uparrow} \to \mathrm{L}^{\uparrow}$ & 14.3 & 1.73 \\
4 & $\underline{\mathrm{L}^{\downarrow} \to \mathrm{L}^{\downarrow} \to \mathrm{L}^{\uparrow} \to \mathrm{L}^{\uparrow}}$ & 11.5 & 1.71 \\
5 & $\mathrm{L}^{\uparrow} \to \mathrm{L}^{\uparrow} \to \mathrm{L}^{\uparrow}$ & 20.0 & 1.68 \\
6 & $\underline{\mathrm{L}^{\downarrow} \to \mathrm{L}^{\uparrow} \to \mathrm{L}^{\uparrow}}$ & 16.3 & 1.67 \\
7 & $\underline{\mathrm{P}^{\downarrow} \to \mathrm{P}^{\downarrow} \to \mathrm{P}^{\downarrow}}$ & 14.8 & 1.56 \\
8 & $\underline{\mathrm{Gap} \to \mathrm{L}^{\uparrow} \to \mathrm{L}^{\uparrow}}$ & 14.8 & 1.50 \\
9 & $\underline{\mathrm{N}^{\downarrow} \to \mathrm{Gap} \to \mathrm{Gap} \to \mathrm{Gap} \to \mathrm{N}^{\uparrow}}$ & 11.1 & 1.48 \\
10 & $\mathrm{L}^{\downarrow} \to \mathrm{Gap} \to \mathrm{Gap} \to \mathrm{Gap} \to \mathrm{L}^{\uparrow}$ & 11.6 & 1.47 \\
11 & $\mathrm{L}^{\downarrow} \to \mathrm{Gap} \to \mathrm{Gap} \to \mathrm{L}^{\uparrow}$ & 13.4 & 1.46 \\
12 & $\mathrm{Gap} \to \mathrm{Gap} \to \mathrm{L}^{\uparrow} \to \mathrm{L}^{\uparrow}$ & 13.1 & 1.46 \\
13 & $\underline{\mathrm{N}^{\downarrow} \to \mathrm{Gap} \to \mathrm{Gap} \to \mathrm{N}^{\uparrow}}$ & 11.8 & 1.45 \\
14 & $\mathrm{P}^{\downarrow} \to \mathrm{Gap} \to \mathrm{P}^{\downarrow}$ & 14.1 & 1.44 \\
15 & $\mathrm{Gap} \to \mathrm{Gap} \to \mathrm{Gap} \to \mathrm{L}^{\uparrow} \to \mathrm{L}^{\uparrow}$ & 11.7 & 1.44 \\
16 & $\underline{\mathrm{N}^{\downarrow} \to \mathrm{Gap} \to \mathrm{Gap} \to \mathrm{Gap} \to \mathrm{N}^{\downarrow}}$ & 23.2 & 1.38 \\
17 & $\mathrm{N}^{\downarrow} \to \mathrm{Gap} \to \mathrm{Gap} \to \mathrm{N}^{\downarrow}$ & 25.3 & 1.34 \\
18 & $\mathrm{L}^{\downarrow} \to \mathrm{L}^{\downarrow} \to \mathrm{Gap} \to \mathrm{Gap} \to \mathrm{L}^{\downarrow}$ & 10.9 & 1.31 \\
19 & $\mathrm{N}^{\downarrow} \to \mathrm{Gap} \to \mathrm{N}^{\downarrow} \to \mathrm{N}^{\downarrow}$ & 11.6 & 1.30 \\
20 & $\underline{\mathrm{L}^{\downarrow} \to \mathrm{Gap} \to \mathrm{Gap} \to \mathrm{L}^{\downarrow}}$ & 19.9 & 1.29 \\
21 & $\mathrm{L}^{\uparrow} \to \mathrm{L}^{\uparrow} \to \mathrm{L}^{\downarrow}$ & 12.0 & 1.29 \\
22 & $\mathrm{L}^{\uparrow} \to \mathrm{L}^{\downarrow} \to \mathrm{L}^{\uparrow}$ & 11.9 & 1.29 \\
23 & $\mathrm{L}^{\downarrow} \to \mathrm{Gap} \to \mathrm{Gap} \to \mathrm{Gap} \to \mathrm{L}^{\downarrow}$ & 16.3 & 1.26 \\
24 & $\mathrm{L}^{\downarrow} \to \mathrm{L}^{\downarrow} \to \mathrm{L}^{\downarrow} \to \mathrm{L}^{\uparrow}$ & 11.5 & 1.25 \\
25 & $\mathrm{L}^{\uparrow} \to \mathrm{L}^{\downarrow} \to \mathrm{L}^{\downarrow} \to \mathrm{L}^{\downarrow}$ & 10.9 & 1.23 \\
26 & $\mathrm{Gap} \to \mathrm{Gap} \to \mathrm{Gap} \to \mathrm{Gap} \to \mathrm{N}^{\uparrow}$ & 14.1 & 1.22 \\
27 & $\mathrm{N}^{\downarrow} \to \mathrm{Gap} \to \mathrm{N}^{\uparrow}$ & 12.6 & 1.21 \\
28 & $\underline{\mathrm{Gap} \to \mathrm{Gap} \to \mathrm{Gap} \to \mathrm{N}^{\uparrow}}$ & 15.0 & 1.20 \\
29 & $\mathrm{N}^{\downarrow} \to \mathrm{Gap} \to \mathrm{N}^{\downarrow}$ & 28.4 & 1.18 \\
30 & $\mathrm{L}^{\downarrow} \to \mathrm{L}^{\downarrow} \to \mathrm{Gap} \to \mathrm{L}^{\downarrow}$ & 14.7 & 1.18 \\
31 & $\mathrm{N}^{\downarrow} \to \mathrm{Gap} \to \mathrm{Gap} \to \mathrm{Gap} \to \mathrm{Gap}$ & 38.9 & 1.17 \\
32 & $\mathrm{N}^{\downarrow} \to \mathrm{Gap} \to \mathrm{Gap} \to \mathrm{Gap}$ & 41.8 & 1.16 \\
33 & $\mathrm{L}^{\downarrow} \to \mathrm{Gap} \to \mathrm{L}^{\downarrow} \to \mathrm{L}^{\downarrow}$ & 13.2 & 1.15 \\
34 & $\mathrm{Gap} \to \mathrm{Gap} \to \mathrm{N}^{\uparrow}$ & 15.8 & 1.15 \\
35 & $\mathrm{Gap} \to \mathrm{Gap} \to \mathrm{Gap} \to \mathrm{Gap}$ & 63.2 & 1.14 \\
36 & $\underline{\mathrm{L}^{\downarrow} \to \mathrm{L}^{\downarrow} \to \mathrm{L}^{\downarrow} \to \mathrm{L}^{\downarrow} \to \mathrm{L}^{\downarrow}}$ & 11.6 & 1.14 \\
37 & $\mathrm{L}^{\downarrow} \to \mathrm{L}^{\downarrow} \to \mathrm{L}^{\uparrow}$ & 16.2 & 1.14 \\
38 & $\mathrm{Gap} \to \mathrm{Gap} \to \mathrm{Gap}$ & 68.7 & 1.13 \\
39 & $\mathrm{P}^{\downarrow} \to \mathrm{P}^{\downarrow} \to \mathrm{Gap}$ & 11.9 & 1.13 \\
40 & $\mathrm{Gap} \to \mathrm{Gap} \to \mathrm{Gap} \to \mathrm{Gap} \to \mathrm{Gap}$ & 57.6 & 1.13 \\
\bottomrule
\end{tabular}
\caption{Top 40 sequential patterns ranked by lift. Symbols denote style and intensity: $\mathrm{L}$ = Liker, $\mathrm{P}$ = Poster, $\mathrm{N}$ = Networker; $\uparrow$ = high intensity, $\downarrow$ = low intensity. Underlined rows are patterns shown in Fig. 5 of the main manuscript.}
\label{tab:prefixspan_lift_top40}
\end{table}

\end{document}